\documentclass[12pt]{article}
\begin{document}

\author{C. Bizdadea\thanks{%
e-mail address: bizdadea@central.ucv.ro}, C. C. Ciob\^\i rc\u {a}\thanks{%
e-mail address: ciobarca@central.ucv.ro}, E. M. Cioroianu\thanks{%
e-mail address: manache@central.ucv.ro}, \and S. O. Saliu\thanks{%
e-mail address: osaliu@central.ucv.ro}, S. C. S\u {a}raru\thanks{%
e-mail address: scsararu@central.ucv.ro} \\
Faculty of Physics, University of Craiova\\
13 A. I. Cuza Str., Craiova RO-1100, Romania}
\title{Four-dimensional couplings among BF and matter theories from BRST cohomology}
\date{}
\maketitle

\begin{abstract}
The local and manifestly covariant Lagrangian interactions in four spacetime
dimensions that can be added to a ``free'' model that describes a generic
matter theory and an abelian BF theory are constructed by means of deforming
the solution to the master equation on behalf of specific cohomological
techniques.

PACS number: 11.10.Ef
\end{abstract}

\section{Introduction}

A big step in the progress of the BRST formalism was its cohomological
understanding \cite{1a}--\cite{1q}, which allowed, among others, a useful
investigation of many interesting aspects related to the perturbative
renormalization problem \cite{4a}--\cite{5}, anomaly-tracking mechanism \cite
{5}--\cite{6}, simultaneous study of local and rigid symmetries of a given
theory \cite{7}, as well as to the reformulation of the construction of
consistent interactions in gauge theories \cite{7a}--\cite{7e} in terms of
the deformation theory \cite{8a}--\cite{8d}, or, actually, in terms of the
deformation of the solution to the master equation. There is a large variety
of models of interest in theoretical physics, which have been investigated
in the light of the deformation of the master equation \cite{9a}--\cite{9za}%
. Some of them focus on the class of BF-like theories \cite{12}. On the one
hand, interacting BF theories are related to Chern-Simons-Witten gravity or
topological two-branes with nonzero three-form. On the other hand, such
theories are important in view of their relationship with Poisson Sigma
Models, which are known to explain interesting aspects of two-dimensional
gravity, including the study of classical solutions \cite{psm1}--\cite{psmn}.

In this paper we construct the local and manifestly covariant interactions
in four spacetime dimensions that can be added to a ``free'' model that
describes a generic matter theory uncoupled to an abelian BF theory \cite{12}%
, by means of deforming the solution to the master equation with the help of
specific cohomological techniques. The field sector of the four-dimensional
BF model consists in one scalar field, two vector fields and one two-form.
Our main result is that we can truly couple the BF fields to the matter ones
in spite of the absence of physical degrees of freedom in the BF model.
Thus, the subject of our paper subscribes to the constant aim of extending
the couplings of more general gauge theories to matter fields.

Our strategy goes as follows. Initially, we generate the ``free'' Lagrangian
BRST symmetry ($s$), which decomposes as the sum between the Koszul-Tate
differential and the exterior longitudinal derivative only. The starting
model is abelian and second-stage reducible, with the reducibility relations
holding off-shell. The only supplementary assumption regards the matter
theory from the perspective of displaying a Cauchy order equal to one.
Nevertheless, this hypothesis is quite natural, since all the usual matter
theories fulfill it. Next, we solve the main equations that govern the
Lagrangian deformation procedure on behalf of the BRST cohomology of the
free theory. In this light, we firstly compute, using specific cohomological
techniques, the first-order deformation of the solution to the master
equation, which lies in the cohomological space of $s$ modulo the exterior
spacetime derivative ($d$) at ghost number zero, $H^{0}\left( s|d\right) $.
The first-order deformation stops at antighost number four and is
parametrized by two arbitrary functions involving only the undifferentiated
scalar field, which we denote by $M\left( \varphi \right) $ and $W\left(
\varphi \right) $. Its consistency demands that $MW=0$, and hence two
distinct situations arise.

\noindent (I) The first one corresponds to $M=0$ and $W$ arbitrary. In this
situation, there appear effective couplings between the matter fields and
the BF ones if the matter theory possesses bosonic one-parameter rigid
symmetries that result in non-trivial conserved currents. From the
inspection of the deformed solution to order one in the coupling constant we
obtain that the interacting theory exhibits the following general features:

\noindent \textit{(i)} all the gauge transformations of the fields from the
BF sector are modified, excepting those of the vector field with original $%
U(1)$ gauge symmetry, which remain $U(1)$-abelian and, moreover, the gauge
transformations of the other vector field include some terms that depend on
the matter fields through the derivative of an arbitrary `background'
potential of the undifferentiated scalar field;

\noindent \textit{(ii)} the matter fields gain gauge transformations, which
at the first order in the deformation parameter are nothing but the gauge
version of the one-parameter rigid symmetry multiplied by the arbitrary
`background' potential;

\noindent \textit{(iii)} the first-order interactions vertices are of two
kinds--- one is responsible for the self-interactions among the BF fields,
while the other couples the matter fields to the $U(1)$-abelian vector field
from the BF model precisely through the conserved current from the matter
theory and the `background' potential (to be called generalized minimal
coupling);

\noindent \textit{(iv)} if the conserved current is not invariant under the
gauge version of the rigid symmetry, then there appear at least second-order
interaction vertices, which truly couple the matter fields to the BF ones,
and further restrictions on the `background' potential are expected.
Otherwise, the deformation comprises first-order couplings;

\noindent \textit{(v)} the deformed gauge algebra is open (unlike the
initial theory, which is abelian);

\noindent \textit{(vi)} the new gauge transformations are second-stage
reducible, like the original ones, but the reducibility relations only hold
on-shell.

\noindent (II) The second situation is described by $W=0$ and $M$ arbitrary.
Then, we find that there are no effective couplings between the BF and the
matter fields. The first-order deformation is also consistent to
higher-orders, which can be taken to vanish. The deformation procedure
simply adds to the free Lagrangian a ``mass''-like term for the two-form.
Related to the gauge transformations, only those of the one-forms change
with respect to the ``free'' model. Accordingly, the gauge algebra becomes
open, and the first-order reducibility relations take place on-shell, while
the second-order ones are not modified with respect to the ``free'' model.

The paper is organized into five sections. Section 2 briefly reviews the
Lagrangian procedure of adding consistent interactions in gauge theories
based on the deformation of the solution to the master equation. In Section
3 we construct the interactions announced in the above, and withdraw the
Lagrangian gauge structure of the coupled models. In section 4 we apply the
theoretical part of the paper to two models of interest, where the role of
the matter fields is played by the complex scalar field, respectively, by
the massive spin 3/2 field. Naturally, we restrict ourselves to the
situation where we can couple the matter fields to the BF ones in a
non-trivial manner. Section 5 ends the paper with the main conclusions.

\section{Deformation of the master equation: a brief review}

We begin with a ``free'' gauge theory, described by a Lagrangian action $%
S_{0}\left[ \Phi ^{\alpha _{0}}\right] $, invariant under some gauge
transformations 
\begin{equation}
\delta _{\epsilon }\Phi ^{\alpha _{0}}=Z_{\;\;\alpha _{1}}^{\alpha
_{0}}\epsilon ^{\alpha _{1}},\;\frac{\delta S_{0}}{\delta \Phi ^{\alpha _{0}}%
}Z_{\;\;\alpha _{1}}^{\alpha _{0}}=0,  \label{bfa2.1}
\end{equation}
and consider the problem of constructing consistent interactions among the
fields $\Phi ^{\alpha _{0}}$ such that the couplings preserve the field
spectrum and the original number of gauge symmetries. This matter is
addressed by means of reformulating the problem of constructing consistent
interactions as a deformation problem of the solution to the master equation
corresponding to the ``free'' theory \cite{8a}, \cite{17and5}. Such a
reformulation is possible due to the fact that the solution to the master
equation contains all the information on the gauge structure of the theory.
If a consistent interacting gauge theory can be constructed, then the
solution $\bar{S}$ to the master equation associated with the ``free''
theory, $\left( \bar{S},\bar{S}\right) =0$, can be deformed into a solution $%
S$, 
\begin{eqnarray}
&&\bar{S}\rightarrow S=\bar{S}+gS_{1}+g^{2}S_{2}+\cdots =  \nonumber \\
&&\bar{S}+g\int d^{D}x\,a+g^{2}\int d^{D}x\,b+\cdots ,  \label{bfa2.2}
\end{eqnarray}
of the master equation for the deformed theory 
\begin{equation}
\left( S,S\right) =0,  \label{bfa2.3}
\end{equation}
such that both the ghost and antifield spectra of the initial theory are
preserved. The symbol $\left( ,\right) $ denotes the antibracket. The
equation (\ref{bfa2.3}) splits, according to the various orders in the
coupling constant (or deformation parameter) $g$, into 
\begin{equation}
\left( \bar{S},\bar{S}\right) =0,  \label{bfa2.4}
\end{equation}
\begin{equation}
2\left( S_{1},\bar{S}\right) =0,  \label{bfa2.5}
\end{equation}
\begin{equation}
2\left( S_{2},\bar{S}\right) +\left( S_{1},S_{1}\right) =0,  \label{bfa2.6}
\end{equation}
\begin{equation}
\left( S_{3},\bar{S}\right) +\left( S_{1},S_{2}\right) =0,  \label{bfa2.7}
\end{equation}
\[
\vdots 
\]

The equation (\ref{bfa2.4}) is fulfilled by hypothesis. The next one
requires that the first-order deformation of the solution to the master
equation, $S_{1}$, is a cocycle of the ``free'' BRST differential $s\cdot
=\left( \cdot ,\bar{S}\right) $. However, only cohomologically non-trivial
solutions to (\ref{bfa2.5}) should be taken into account, as the BRST-exact
ones can be eliminated by a (in general non-linear) field redefinition. This
means that $S_{1}$ pertains to the ghost number zero cohomological space of $%
s$, $H^{0}\left( s\right) $, which is generically nonempty due to its
isomorphism to the space of physical observables of the ``free'' theory. It
has been shown in \cite{8a}, \cite{17and5} (on behalf of the triviality of
the antibracket map in the cohomology of the BRST differential) that there
are no obstructions in finding solutions to the remaining equations ((\ref
{bfa2.6}--\ref{bfa2.7}), etc.). However, the resulting interactions may be
nonlocal, and there might even appear obstructions if one insists on their
locality. The analysis of these obstructions can be done with the help of
cohomological techniques. As it will be seen below, all the interactions in
the case of the model under study turn out to be local.

\section{Couplings among a BF theory and matter fields from BRST cohomology}

In this section we determine the local and manifestly covariant Lagrangian
interactions that can be added to a ``free'' theory that describes a generic
matter theory plus a topological model of BF-type in four spacetime
dimensions. This is done by means of solving the deformation equations (\ref
{bfa2.5}--\ref{bfa2.7}), etc., by means of specific cohomological
techniques. The interacting theory and its gauge structure are deduced from
the analysis of the deformed solution to the master equation that is
consistent to all orders in the deformation parameter.

\subsection{Free BRST differential}

We start from a ``free'' four-dimensional theory whose Lagrangian action is
written as the sum between the action for a matter theory and the action for
a topological BF theory involving one scalar field, two one-forms and one
two-form 
\begin{eqnarray}
&&S_{0}[A^{\mu },H^{\mu },\varphi ,B^{\mu \nu },y^{i}]=\int d^{4}x\left(
H_{\mu }\partial ^{\mu }\varphi +\frac{1}{2}B^{\mu \nu }\partial _{[\mu
}A_{\nu ]}\right.  \nonumber \\
&&\left. +\mathcal{L}_{0}^{\mathrm{matt}}\left( y^{i},\partial _{\mu
}y^{i},\cdots \right) \right) \equiv \int d^{4}x\left( \mathcal{L}_{0}^{%
\mathrm{BF}}+\mathcal{L}_{0}^{\mathrm{matt}}\right) ,  \label{bfa1}
\end{eqnarray}
and suppose that the matter theory displays no non-trivial gauge symmetries.
We work with a Minkowski-flat metric tensor of `mostly minus' signature $%
g^{\mu \nu }=g_{\mu \nu }=\left( +---\right) $. The Grassmann parity of a
given matter field $y^{i}$ will be denoted by $\varepsilon _{i}$. In
addition, we make the assumption that $\mathcal{L}_{0}^{\mathrm{matt}}$
decomposes like 
\begin{equation}
\mathcal{L}_{0}^{\mathrm{matt}}=\mathcal{L}_{0}^{\mathrm{free}}+\mathcal{L}%
_{0}^{\mathrm{int}},  \label{decmatter}
\end{equation}
where its free part, $\mathcal{L}_{0}^{\mathrm{free}}$, is quadratic in the
fields $y^{i}$ and of maximum order two in their spacetime derivatives,
while its interacting part, $\mathcal{L}_{0}^{\mathrm{int}}$, if present, is
no more than a polynomial in the undifferentiated fields. This assumption
combined with the absence of gauge invariance for $\mathcal{L}_{0}^{\mathrm{%
matt}}$ leads to the result that the matter sector is described by a
so-called normal theory, of Cauchy order equal to one. This is not a
restrictive condition, but merely a natural one, since all usual matter
theories (like, for instance, those of spin 0, 1/2 or 3/2) satisfy it. In
turn, it enables one to control in a consistent manner the local cohomology
of the BRST differential associated with $S_{0}$.

Action (\ref{bfa1}) is found invariant under the gauge transformations 
\begin{equation}
\delta _{\epsilon }A^{\mu }=\partial ^{\mu }\epsilon ,\;\delta _{\epsilon
}H^{\mu }=2\partial _{\nu }\epsilon ^{\mu \nu },\;\delta _{\epsilon }\varphi
=0,\;\delta _{\epsilon }B^{\mu \nu }=-3\partial _{\rho }\epsilon ^{\mu \nu
\rho },\;\delta _{\epsilon }y^{i}=0,  \label{bfa2}
\end{equation}
where the gauge parameters $\epsilon $, $\epsilon ^{\mu \nu }$ and $\epsilon
^{\mu \nu \rho }$ are bosonic, with $\epsilon ^{\mu \nu }$ and $\epsilon
^{\mu \nu \rho }$ completely antisymmetric. From (\ref{bfa2}), we read the
non-vanishing gauge generators 
\begin{equation}
(Z_{(A)}^{\mu })(x,x^{\prime })=\partial _{x}^{\mu }\delta ^{4}(x-x^{\prime
}),\;(Z_{(H)}^{\mu })_{\alpha \beta }(x,x^{\prime })=-\partial _{\left[
\alpha \right. }^{x}\delta _{\left. \beta \right] }^{\mu }\delta
^{4}(x-x^{\prime }),  \label{bfa2a}
\end{equation}
\begin{equation}
(Z_{(B)}^{\mu \nu })_{\alpha \beta \gamma }(x,x^{\prime })=-\frac{1}{2}%
\partial _{\left[ \alpha \right. }^{x}\delta _{\beta }^{\mu }\delta _{\left.
\gamma \right] }^{\nu }\delta ^{4}(x-x^{\prime }),  \label{bfa2b}
\end{equation}
where we put an extra lower index ($(A)$, $(H)$, etc.) in order to indicate
with what field is a certain gauge generator associated. Everywhere in this
paper we use the convention that the notation $\left[ \alpha \beta \cdots
\gamma \right] $ signifies complete antisymmetry with respect to the Lorentz
indices between brackets, with no additional numerical factor. The above
gauge transformations are abelian and off-shell second-order reducible. More
precisely, the gauge generators of the one-form $H^{\mu }$ are second-order
reducible, with the first-, respectively, second-order reducibility
functions 
\begin{equation}
(Z_{1}^{\alpha \beta })_{\mu ^{\prime }\nu ^{\prime }\rho ^{\prime
}}(x,x^{\prime })=-\frac{1}{2}\partial _{\left[ \mu ^{\prime }\right.
}^{x}\delta _{\nu ^{\prime }}^{\alpha }\delta _{\left. \rho ^{\prime
}\right] }^{\beta }\delta ^{4}(x-x^{\prime }),  \label{bfa3}
\end{equation}
\begin{equation}
(Z_{2}^{\mu ^{\prime }\nu ^{\prime }\rho ^{\prime }})_{\alpha ^{\prime
}\beta ^{\prime }\gamma ^{\prime }\delta ^{\prime }}(x,x^{\prime })=-\frac{1%
}{6}\partial _{\left[ \alpha ^{\prime }\right. }^{x}\delta _{\beta ^{\prime
}}^{\mu ^{\prime }}\delta _{\gamma ^{\prime }}^{\nu ^{\prime }}\delta
_{\left. \delta ^{\prime }\right] }^{\rho ^{\prime }}\delta ^{4}(x-x^{\prime
}),  \label{bfa4}
\end{equation}
while the gauge generators of the two-form $B^{\mu \nu }$ are first-order
reducible, with the reducibility functions 
\begin{equation}
(Z_{1}^{\alpha \beta \gamma })_{\mu ^{\prime }\nu ^{\prime }\rho ^{\prime
}\lambda ^{\prime }}(x,x^{\prime })=-\frac{1}{6}\partial _{\left[ \mu
^{\prime }\right. }^{x}\delta _{\nu ^{\prime }}^{\alpha }\delta _{\rho
^{\prime }}^{\beta }\delta _{\left. \lambda ^{\prime }\right] }^{\gamma
}\delta ^{4}(x-x^{\prime }),  \label{bfa5}
\end{equation}
such that the concrete form of the first- and second-order reducibility
relations written in condensed De Witt notations are expressed by 
\begin{equation}
(Z_{(H)}^{\mu })_{\alpha \beta }(Z_{1}^{\alpha \beta })_{\mu ^{\prime }\nu
^{\prime }\rho ^{\prime }}=0,\;(Z_{(B)}^{\mu \nu })_{\alpha \beta \gamma
}(Z_{1}^{\alpha \beta \gamma })_{\mu ^{\prime }\nu ^{\prime }\rho ^{\prime
}\lambda ^{\prime }}=0,  \label{bfa5a}
\end{equation}
respectively, 
\begin{equation}
(Z_{1}^{\alpha \beta })_{\mu ^{\prime }\nu ^{\prime }\rho ^{\prime
}}(Z_{2}^{\mu ^{\prime }\nu ^{\prime }\rho ^{\prime }})_{\alpha ^{\prime
}\beta ^{\prime }\gamma ^{\prime }\delta ^{\prime }}=0.  \label{bfa5b}
\end{equation}
We observe that the BF theory alone is a usual linear gauge theory (its
field equations are linear in the fields and first-order in their spacetime
derivatives), whose generating set of gauge transformations is second-order
reducible, such that we can define in a consistent manner its Cauchy order,
which is found equal to four.

In order to construct the BRST symmetry of this ``free'' theory, we
introduce the field/ghost and antifield spectra 
\begin{equation}
\Phi ^{\alpha _{0}}=\left( A^{\mu },H^{\mu },\varphi ,B^{\mu \nu
},y^{i}\right) ,\;\Phi _{\alpha _{0}}^{*}=\left( A_{\mu }^{*},H_{\mu
}^{*},\varphi ^{*},B_{\mu \nu }^{*},y_{i}^{*}\right) ,  \label{bfa6}
\end{equation}
\begin{equation}
\eta ^{\alpha _{1}}=\left( \eta ,C^{\mu \nu },\eta ^{\mu \nu \rho }\right)
,\;\eta _{\alpha _{1}}^{*}=\left( \eta ^{*},C_{\mu \nu }^{*},\eta _{\mu \nu
\rho }^{*}\right) ,  \label{bfa7}
\end{equation}
\begin{equation}
\eta ^{\alpha _{2}}=\left( C^{\mu \nu \rho },\eta ^{\mu \nu \rho \lambda
}\right) ,\;\eta _{\alpha _{2}}^{*}=\left( C_{\mu \nu \rho }^{*},\eta _{\mu
\nu \rho \lambda }^{*}\right) ,  \label{bfa8}
\end{equation}
\begin{equation}
\eta ^{\alpha _{3}}=C^{\mu \nu \rho \lambda },\;\eta _{\alpha
_{3}}^{*}=C_{\mu \nu \rho \lambda }^{*}.  \label{bfa9}
\end{equation}
The fermionic ghosts $\eta ^{\alpha _{1}}$ respectively correspond to the
bosonic gauge parameters $\epsilon ^{\alpha _{1}}=\left( \epsilon ,\epsilon
^{\mu \nu },\epsilon ^{\mu \nu \rho }\right) $, the bosonic ghosts for
ghosts $\eta ^{\alpha _{2}}$ are due to the first-order reducibility
relations (\ref{bfa5a}), while the fermionic ghosts for ghosts for ghosts $%
\eta ^{\alpha _{3}}$ are required by the second-order reducibility relations
(\ref{bfa5b}). The star variables represent the antifields of the
corresponding fields/ghosts. Their Grassmann parities are obtained via the
usual rule 
\[
\varepsilon \left( \chi ^{*}\right) =\left( \varepsilon \left( \chi \right)
+1\right) \mathrm{mod}\;2, 
\]
where we employed the notations 
\begin{equation}
\chi =\left( \Phi ^{\alpha _{0}},\eta ^{\alpha _{1}},\eta ^{\alpha
_{2}},\eta ^{\alpha _{3}}\right) ,\;\chi ^{*}=\left( \Phi _{\alpha
_{0}}^{*},\eta _{\alpha _{1}}^{*},\eta _{\alpha _{2}}^{*},\eta _{\alpha
_{3}}^{*}\right) .  \label{notat}
\end{equation}

Since both the gauge generators and the reducibility functions are
field-independent, it follows that the BRST differential reduces to 
\begin{equation}
s=\delta +\gamma ,  \label{sfree}
\end{equation}
where $\delta $ is the Koszul-Tate differential, and $\gamma $ means the
exterior longitudinal derivative. The Koszul-Tate differential is graded in
terms of the antighost number ($\mathrm{agh}$, $\mathrm{agh}\left( \delta
\right) =-1$, $\mathrm{agh}\left( \gamma \right) =0$) and enforces a
resolution of the algebra of smooth functions defined on the stationary
surface of field equations for action (\ref{bfa1}), $C^{\infty }\left(
\Sigma \right) $, $\Sigma :\delta S_{0}/\delta \Phi ^{\alpha _{0}}=0$. The
exterior longitudinal derivative is graded in terms of the pure ghost number
($\mathrm{pgh}$, $\mathrm{pgh}\left( \gamma \right) =1$, $\mathrm{pgh}\left(
\delta \right) =0$) and is correlated with the original gauge symmetry via
its cohomology at pure ghost number zero computed in $C^{\infty }\left(
\Sigma \right) $, which is isomorphic to the algebra of physical observables
for the ``free'' theory. The two degrees of the generators (\ref{bfa6}--\ref
{bfa9}) from the BRST complex are valued like 
\begin{equation}
\mathrm{pgh}\left( \Phi ^{\alpha _{0}}\right) =0,\;\mathrm{pgh}\left( \eta
^{\alpha _{1}}\right) =1,\;\mathrm{pgh}\left( \eta ^{\alpha _{2}}\right)
=2,\;\mathrm{pgh}\left( \eta ^{\alpha _{3}}\right) =3,  \label{bfa10}
\end{equation}
\begin{equation}
\mathrm{pgh}\left( \Phi _{\alpha _{0}}^{*}\right) =\mathrm{pgh}\left( \eta
_{\alpha _{1}}^{*}\right) =\mathrm{pgh}\left( \eta _{\alpha _{2}}^{*}\right)
=\mathrm{pgh}\left( \eta _{\alpha _{3}}^{*}\right) =0,  \label{bfa12}
\end{equation}
\begin{equation}
\mathrm{agh}\left( \Phi ^{\alpha _{0}}\right) =\mathrm{agh}\left( \eta
^{\alpha _{1}}\right) =\mathrm{agh}\left( \eta ^{\alpha _{2}}\right) =%
\mathrm{agh}\left( \eta ^{\alpha _{3}}\right) =0,  \label{bfa11}
\end{equation}
\begin{equation}
\mathrm{agh}\left( \Phi _{\alpha _{0}}^{*}\right) =1,\;\mathrm{agh}\left(
\eta _{\alpha _{1}}^{*}\right) =2,\;\mathrm{agh}\left( \eta _{\alpha
_{2}}^{*}\right) =3,\;\mathrm{agh}\left( \eta _{\alpha _{3}}^{*}\right) =4,
\label{bfa13}
\end{equation}
where the actions of $\delta $ and $\gamma $ on them read as 
\begin{equation}
\delta \Phi ^{\alpha _{0}}=\delta \eta ^{\alpha _{1}}=\delta \eta ^{\alpha
_{2}}=\delta \eta ^{\alpha _{3}}=0,  \label{bfa15}
\end{equation}
\begin{equation}
\delta A_{\mu }^{*}=\partial ^{\nu }B_{\nu \mu },\;\delta H_{\mu
}^{*}=-\partial _{\mu }\varphi ,\;\delta \varphi ^{*}=\partial ^{\mu }H_{\mu
},\;\delta B_{\mu \nu }^{*}=-\frac{1}{2}\partial _{[\mu }A_{\nu ]},
\label{bfa16}
\end{equation}
\begin{equation}
\delta y_{i}^{*}=-\frac{\delta ^{L}\mathcal{L}_{0}^{\mathrm{matt}}}{\delta
y^{i}},\;\delta \eta ^{*}=-\partial ^{\mu }A_{\mu }^{*},\;\delta C_{\mu \nu
}^{*}=\partial _{[\mu }H_{\nu ]}^{*},  \label{bfa17}
\end{equation}
\begin{equation}
\delta \eta _{\mu \nu \rho }^{*}=\partial _{[\mu }B_{\nu \rho
]}^{*},\;\delta C_{\mu \nu \rho }^{*}=-\partial _{[\mu }C_{\nu \rho ]}^{*},
\label{bfa18}
\end{equation}
\begin{equation}
\delta \eta _{\mu \nu \rho \lambda }^{*}=-\partial _{[\mu }\eta _{\nu \rho
\lambda ]}^{*},\;\delta C_{\mu \nu \rho \lambda }^{*}=\partial _{[\mu
}C_{\nu \rho \lambda ]}^{*},  \label{bfa19}
\end{equation}
\begin{equation}
\gamma \Phi _{\alpha _{0}}^{*}=\gamma \eta _{\alpha _{1}}^{*}=\gamma \eta
_{\alpha _{2}}^{*}=\gamma \eta _{\alpha _{3}}^{*}=0,  \label{bfa20}
\end{equation}
\begin{equation}
\gamma A^{\mu }=\partial ^{\mu }\eta ,\;\gamma H^{\mu }=2\partial _{\nu
}C^{\mu \nu },\;\gamma B^{\mu \nu }=-3\partial _{\rho }\eta ^{\mu \nu \rho
},\;\gamma \varphi =\gamma y^{i}=0,  \label{bfa21}
\end{equation}
\begin{equation}
\gamma \eta =0,\;\gamma C^{\mu \nu }=-3\partial _{\rho }C^{\mu \nu \rho
},\;\gamma \eta ^{\mu \nu \rho }=4\partial _{\lambda }\eta ^{\mu \nu \rho
\lambda },  \label{bfa22}
\end{equation}
\begin{equation}
\gamma C^{\mu \nu \rho }=4\partial _{\lambda }C^{\mu \nu \rho \lambda
},\;\gamma \eta ^{\mu \nu \rho \lambda }=\gamma C^{\mu \nu \rho \lambda }=0.
\label{bfa23}
\end{equation}

The overall degree of the BRST complex is named ghost number ($\mathrm{gh}$)
and is defined like the difference between the pure ghost number and the
antighost number, such that $\mathrm{gh}\left( s\right) =1$. The BRST
symmetry admits a canonical action $s\cdot =\left( \cdot ,\bar{S}\right) $,
where its canonical generator ($\mathrm{gh}\left( \bar{S}\right) =0$, $%
\varepsilon \left( \bar{S}\right) =0$) satisfies the classical master
equation $\left( \bar{S},\bar{S}\right) =0$. In the case of the ``free''
theory under discussion, the solution to the master equation takes the form 
\begin{eqnarray}
&&\bar{S}=S_{0}+\int d^{4}x\left( A_{\mu }^{*}\partial ^{\mu }\eta +2H_{\mu
}^{*}\partial _{\nu }C^{\mu \nu }-3B_{\mu \nu }^{*}\partial _{\rho }\eta
^{\mu \nu \rho }\right.  \nonumber \\
&&\left. -3C_{\mu \nu }^{*}\partial _{\rho }C^{\mu \nu \rho }+4\eta _{\mu
\nu \rho }^{*}\partial _{\lambda }\eta ^{\mu \nu \rho \lambda }+4C_{\mu \nu
\rho }^{*}\partial _{\lambda }C^{\mu \nu \rho \lambda }\right) .
\label{bfa14}
\end{eqnarray}
The solution to the master equation encodes all the information on the gauge
structure of a given theory. We remark that in our case the solution (\ref
{bfa14}) to the master equation breaks into terms with antighost numbers
ranging from zero to three. The piece with antighost number zero is nothing
but the Lagrangian action (\ref{bfa1}), while the elements of antighost
number one include the gauge generators (\ref{bfa2a}--\ref{bfa2b}). If the
gauge algebra is non-abelian, then there appear terms linear in the
antighost number two antifields and quadratic in the pure ghost number one
ghosts. The absence of such terms in our case reflects that the gauge
transformations are abelian. The terms from (\ref{bfa14}) with higher
antighost number give us information on the reducibility functions (\ref
{bfa3}--\ref{bfa5}). If the reducibility relations held on-shell, then there
would appear components linear in the ghosts for ghosts (ghosts of pure
ghost number strictly greater than one) and quadratic in the various
antifields. Such pieces are not present in (\ref{bfa14}), since the
reducibility relations hold off-shell. Other possible components in the
solution to the master equation offer information on the higher-order
structure functions related to the tensor gauge structure of the theory.
There are no such terms in (\ref{bfa14}), as a consequence of the fact that
all higher-order structure functions vanish for the theory under study.

\subsection{First-order deformations}

Initially, we approach the first-order deformation of the solution to the
master equation, described by the equation (\ref{bfa2.5}). In local form, it
becomes 
\begin{equation}
sa=\partial _{\mu }n^{\mu },  \label{bfa24}
\end{equation}
for some local $n^{\mu }$ (where we maintained the notations from (\ref
{bfa2.2})), so it requires that $a$ is a $s$-cocycle modulo $d$. In order to
solve this equation, we develop $a$ according to the antighost number, 
\begin{equation}
a=a_{0}+a_{1}+\cdots +a_{I},\;\mathrm{agh}\left( a_{J}\right) =J,\;\mathrm{gh%
}\left( a_{J}\right) =0,\;\varepsilon \left( a_{J}\right) =0,\;J=\overline{%
0,I}.  \label{bfa25}
\end{equation}
The number of terms in the expansion (\ref{bfa25}) is finite and it can be
shown that we can take the last term in $a$ to be annihilated by $\gamma $, 
\begin{equation}
\gamma a_{I}=0.  \label{bfa26}
\end{equation}
The fact that the free BRST differential is just the sum of $\delta $ and $%
\gamma $ (see (\ref{sfree})), and so it is not an infinite formal series of
derivations with arbitrarily high antighost number (as it can a priori occur
for an arbitrary gauge system with an open gauge algebra), argues that the
non-integrated density of the first-order deformation $a$ can be assumed,
without loss of generality, to have a bounded antighost number, say $I$. A
more rigorous proof of the validity of this result can be obtained by
following the line from \cite{9d} under the sole assumption that the
first-order deformation of the Lagrangian $a_{0}$ has a finite (but
otherwise arbitrary) derivative order. More precisely, one can prove that
the analogue of Theorem 3.1 therein is valid for the ``free'' model
considered in this paper. This can be done by introducing an even derivation 
$K=N_{\partial }+A$, where $N_{\partial }$ is the operator counting the
derivatives of all the variables and $A$ gives various weights to the
antifields and ghosts, such that both $\gamma $ and $\delta $ have only
components of non positive $K$-degree. In our case it is possible to define $%
A$ such that $\gamma $ and $\delta $ actually reduce to their components of
zero $K$-degree. The result that the term of highest antighost number in (%
\ref{bfa25}) $a_{I}$ can be taken to satisfy (\ref{bfa26}) rather than the
obvious equation $\gamma a_{I}=\partial _{\mu }u^{\mu }$ that follows from (%
\ref{bfa24}) (with $a$ replaced by (\ref{bfa25})) projected on antighost
number $I$ can be proven like in \cite{9u} (see Section 3 and also the
Appendix A.1 therein) and is related to the triviality of the invariant
cohomology of the exterior spacetime differential $d$ in form degree less
than four and in strictly positive antighost number.

Consequently, we need to compute the cohomology of $\gamma $, $H\left(
\gamma \right) $, in order to determine the component of highest antighost
number in $a$. From (\ref{bfa20}--\ref{bfa23}) it is simple to see that $%
H\left( \gamma \right) $ is spanned by $F^{\mu \nu }=\partial ^{\left[ \mu
\right. }A^{\left. \nu \right] }$, $\partial _{\mu }H^{\mu }$, $\varphi $, $%
\partial _{\mu }B^{\mu \nu }$, $y^{i}$, and the antifields $\chi ^{*}$ from (%
\ref{notat}), by their spacetime derivatives, as well as by the
undifferentiated ghosts $\eta ^{A_{1}}=\left( \eta ,\eta ^{\mu \nu \rho
\lambda },C_{\mu \nu \rho \lambda }\right) $. (The derivatives of the ghosts 
$\eta ^{A_{1}}$ are removed from $H\left( \gamma \right) $ since they are $%
\gamma $-exact, in agreement with the first relation in (\ref{bfa21}), the
last formula in (\ref{bfa22}), respectively, the first definition from (\ref
{bfa23}).) If we denote by $e^{M}\left( \eta ^{A_{1}}\right) $ the elements
with pure ghost number $M$ of a basis in the space of the polynomials in the
ghosts $\eta ^{A_{1}}$, it follows that the general solution to the equation
(\ref{bfa26}) takes the form 
\begin{equation}
a_{I}=\mu _{I}\left( \left[ F^{\mu \nu }\right] ,\left[ \partial _{\mu
}H^{\mu }\right] ,\left[ \varphi \right] ,\left[ \partial _{\mu }B^{\mu \nu
}\right] ,\left[ y^{i}\right] ,\left[ \chi ^{*}\right] \right) e^{I}\left(
\eta ^{A_{1}}\right) ,  \label{bfa27}
\end{equation}
where $\mathrm{agh}\left( \mu _{I}\right) =I$ and $\mathrm{pgh}\left(
e^{I}\right) =I$. The notation $f\left( \left[ q\right] \right) $ means that 
$f$ depends on $q$ and its spacetime derivatives up to a finite order. The
equation (\ref{bfa24}) projected on antighost number $\left( I-1\right) $
becomes 
\begin{equation}
\delta a_{I}+\gamma a_{I-1}=\partial ^{\mu }\stackrel{(I-1)}{m}_{\mu }.
\label{bfa28}
\end{equation}
Replacing (\ref{bfa27}) in (\ref{bfa28}), it follows that the last equation
possesses solutions with respect to $a_{I-1}$ if the coefficients $\mu _{I}$
pertain to the homological space $H_{I}\left( \delta |d\right) $, i.e., $%
\delta \mu _{I}=\partial _{\mu }l_{I-1}^{\mu }$. In order to analyze the
local homology of the Koszul-Tate differential, $H\left( \delta |d\right) $,
we observe that the form (\ref{bfa1}) of the ``free'' Lagrangian action
together with the definitions (\ref{bfa15}--\ref{bfa19}) enable us to
analyse $H_{J}\left( \delta |d\right) $ in terms of the local homologies $%
H_{J}^{\mathrm{matt}}\left( \delta |d\right) $ and $H_{J}^{\mathrm{BF}%
}\left( \delta |d\right) $, where the last local homologies refer to the
Koszul-Tate operator that acts non-trivially only in the matter sector,
respectively, only in the BF one\footnote{%
Indeed, we can decompose $\delta $ like $\delta =\delta ^{\mathrm{matt}%
}+\delta ^{\mathrm{BF}}$, where $\delta ^{\mathrm{matt}}\left( \mathrm{%
matter\;variables}\right) =\delta \left( \mathrm{matter\;variables}\right) $
and $\delta ^{\mathrm{matt}}\left( \mathrm{BF\;variables}\right) =0$,
respectively, $\delta ^{\mathrm{BF}}\left( \mathrm{matter\;variables}\right)
=0$ and $\delta ^{\mathrm{BF}}\left( \mathrm{BF\;variables}\right) =\delta
\left( \mathrm{BF\;variables}\right) $. According to this decomposition, $H^{%
\mathrm{matt}}\left( \delta |d\right) $ and $H^{\mathrm{BF}}\left( \delta
|d\right) $ must be understood only as some more suggestive notations for $%
H\left( \delta ^{\mathrm{matt}}|d\right) $ and $H\left( \delta ^{\mathrm{BF}%
}|d\right) $ respectively.}. In the light of the general results from \cite
{gen1}--\cite{gen2}, the assumption on the behavior of the matter theory,
more precisely on its Cauchy order, which was supposed to be equal to one,
combined with the fact that the BF component is separately described by a
linear theory of Cauchy order equal to four, guarantee that 
\begin{equation}
H_{J}^{\mathrm{matt}}\left( \delta |d\right) =0,\;J>1.  \label{hmat}
\end{equation}
\begin{equation}
H_{J}^{\mathrm{BF}}\left( \delta |d\right) =0,\;J>4,  \label{hbf}
\end{equation}
such that we conclude that $H_{J}\left( \delta |d\right) =0$ for $J>4$, and,
moreover, $H_{J}\left( \delta |d\right) =H_{J}^{\mathrm{BF}}\left( \delta
|d\right) $ for $J=2,3,4$. However, in principle the representatives of $%
H_{1}\left( \delta |d\right) $ cannot be written like sums between
representatives of $H_{1}^{\mathrm{matt}}\left( \delta |d\right) $ and of $%
H_{1}^{\mathrm{BF}}\left( \delta |d\right) $, such that their study deserves
special attention. Nevertheless, we can assume that the first-order
deformation stops at antighost number four ($I=4$) 
\begin{equation}
a=a_{0}+a_{1}+a_{2}+a_{3}+a_{4},  \label{bfa29}
\end{equation}
where $a_{4}$ is of the form (\ref{bfa27}), with $\mu _{4}$ from $%
H_{4}\left( \delta |d\right) =H_{4}^{\mathrm{BF}}\left( \delta |d\right) $.
This means that the matter theory cannot be involved with the components $%
\left( a_{J}\right) _{J=2,3,4}$. It acts non-trivially only at antighost
number one, where its contribution to $a_{1}$ can be introduced precisely
via the `homogeneous' equation $\gamma \bar{a}_{1}=0$. This already
eliminates the dependence on $y^{i}$, $y_{i}^{*}$ and their spacetime
derivatives from (\ref{bfa27}) for $I\rightarrow J=2,3,4$.

By direct computation, we infer that the most general representative of $%
H_{4}^{\mathrm{BF}}\left( \delta |d\right) $ can be taken of the type 
\begin{eqnarray}
&&\left( \mu _{4}\right) _{\mu \nu \rho \lambda }=\frac{\delta W}{\delta
\varphi }C_{\mu \nu \rho \lambda }^{*}+\frac{\delta ^{2}W}{\delta \varphi
^{2}}H_{\left[ \mu \right. }^{*}C_{\left. \nu \rho \lambda \right] }^{*}+%
\frac{\delta ^{2}W}{\delta \varphi ^{2}}C_{\left[ \mu \nu \right.
}^{*}C_{\left. \rho \lambda \right] }^{*}  \nonumber \\
&&+\frac{\delta ^{3}W}{\delta \varphi ^{3}}H_{\left[ \mu \right. }^{*}H_{\nu
}^{*}C_{\left. \rho \lambda \right] }^{*}+\frac{\delta ^{4}W}{\delta \varphi
^{4}}H_{\mu }^{*}H_{\nu }^{*}H_{\rho }^{*}H_{\lambda }^{*},  \label{bfa30}
\end{eqnarray}
with $W=W(\varphi )$ an arbitrary function depending on the undifferentiated
scalar field. On the other hand, the elements of pure ghost number equal to
four of the basis in the ghosts $\eta ^{A_{1}}$ are 
\begin{equation}
\eta C^{\mu \nu \rho \lambda },\eta ^{\alpha \beta \gamma \delta }\eta
^{\alpha ^{\prime }\beta ^{\prime }\gamma ^{\prime }\delta ^{\prime }}.
\label{bfa30a}
\end{equation}
In order to couple (\ref{bfa30}) to the second element in (\ref{bfa30a})
like in (\ref{bfa27}) we need some completely antisymmetric constants,
which, by covariance arguments, can only be proportional with the completely
antisymmetric four-dimensional symbol, $\varepsilon _{\alpha \beta \gamma
\delta }$. Apparently, there are several possibilities to realize such
couplings. However, all these possibilities lead to the same result, such
that there is in fact a single independent manner in which (\ref{bfa30}) can
be ``glued'' to the latter basis elements in (\ref{bfa30a}), and thus the
most general (manifestly covariant) form of the last representative from the
expansion (\ref{bfa29}) will be 
\begin{eqnarray}
&&a_{4}=\left( \frac{\delta W}{\delta \varphi }C_{\mu \nu \rho \lambda }^{*}+%
\frac{\delta ^{2}W}{\delta \varphi ^{2}}H_{\left[ \mu \right. }^{*}C_{\left.
\nu \rho \lambda \right] }^{*}+\frac{\delta ^{2}W}{\delta \varphi ^{2}}%
C_{\left[ \mu \nu \right. }^{*}C_{\left. \rho \lambda \right] }^{*}\right. 
\nonumber \\
&&\left. +\frac{\delta ^{3}W}{\delta \varphi ^{3}}H_{\left[ \mu \right.
}^{*}H_{\nu }^{*}C_{\left. \rho \lambda \right] }^{*}+\frac{\delta ^{4}W}{%
\delta \varphi ^{4}}H_{\mu }^{*}H_{\nu }^{*}H_{\rho }^{*}H_{\lambda
}^{*}\right) \eta C^{\mu \nu \rho \lambda }  \nonumber \\
&&+\frac{1}{2}\left( \frac{\delta M}{\delta \varphi }C_{\mu \nu \rho \lambda
}^{*}+\frac{\delta ^{2}M}{\delta \varphi ^{2}}H_{\left[ \mu \right.
}^{*}C_{\left. \nu \rho \lambda \right] }^{*}+\frac{\delta ^{2}M}{\delta
\varphi ^{2}}C_{\left[ \mu \nu \right. }^{*}C_{\left. \rho \lambda \right]
}^{*}\right.  \nonumber \\
&&\left. +\frac{\delta ^{3}M}{\delta \varphi ^{3}}H_{\left[ \mu \right.
}^{*}H_{\nu }^{*}C_{\left. \rho \lambda \right] }^{*}+\frac{\delta ^{4}M}{%
\delta \varphi ^{4}}H_{\mu }^{*}H_{\nu }^{*}H_{\rho }^{*}H_{\lambda
}^{*}\right) \eta ^{\mu \nu \rho \lambda }\varepsilon _{\alpha \beta \gamma
\delta }\eta ^{\alpha \beta \gamma \delta },  \label{bfa31}
\end{eqnarray}
where the numerical factor $1/2$ in the second term was taken for
convenience, and the functions $W$ and $M$ are two arbitrary functions of
the undifferentiated scalar field. By computing the action of $\delta $ on $%
a_{4}$ and by taking into account the relations (\ref{bfa20}--\ref{bfa23}),
it follows that the solution of the equation (\ref{bfa28}) for $I=4$ is
precisely given by 
\begin{eqnarray}
&&a_{3}=-\left( \frac{\delta W}{\delta \varphi }C_{\nu \rho \lambda }^{*}+%
\frac{\delta ^{2}W}{\delta \varphi ^{2}}H_{\left[ \nu \right. }^{*}C_{\left.
\rho \lambda \right] }^{*}+\frac{\delta ^{3}W}{\delta \varphi ^{3}}H_{\nu
}^{*}H_{\rho }^{*}H_{\lambda }^{*}\right) \times  \nonumber \\
&&\times \left( 4A_{\mu }C^{\mu \nu \rho \lambda }+\eta C^{\nu \rho \lambda
}\right) +2W\eta _{\mu \nu \rho \lambda }^{*}C^{\mu \nu \rho \lambda }-8%
\frac{\delta W}{\delta \varphi }H_{\lambda }^{*}\eta _{\mu \nu \rho
}^{*}C^{\mu \nu \rho \lambda }  \nonumber \\
&&+12\left( \frac{\delta W}{\delta \varphi }C_{\rho \lambda }^{*}+\frac{%
\delta ^{2}W}{\delta \varphi ^{2}}H_{\rho }^{*}H_{\lambda }^{*}\right)
B_{\mu \nu }^{*}C^{\mu \nu \rho \lambda }-\left( \frac{\delta M}{\delta
\varphi }C_{\nu \rho \lambda }^{*}\right.  \nonumber \\
&&\left. +\frac{\delta ^{2}M}{\delta \varphi ^{2}}H_{\left[ \nu \right.
}^{*}C_{\left. \rho \lambda \right] }^{*}+\frac{\delta ^{3}M}{\delta \varphi
^{3}}H_{\nu }^{*}H_{\rho }^{*}H_{\lambda }^{*}\right) \eta ^{\nu \rho
\lambda }\varepsilon _{\alpha \beta \gamma \delta }\eta ^{\alpha \beta
\gamma \delta }.  \label{bfa33}
\end{eqnarray}
By means of the equation (\ref{bfa24}) projected on antighost number two 
\begin{equation}
\delta a_{3}+\gamma a_{2}=\partial ^{\mu }\stackrel{(2)}{m}_{\mu },
\label{bfa34}
\end{equation}
the solution (\ref{bfa33}) and the definitions (\ref{bfa20}--\ref{bfa23})
lead to 
\begin{eqnarray}
&&a_{2}=\left( \frac{\delta W}{\delta \varphi }C_{\mu \nu }^{*}+\frac{\delta
^{2}W}{\delta \varphi ^{2}}H_{\mu }^{*}H_{\nu }^{*}\right) \left(
-3A_{\lambda }C^{\mu \nu \lambda }+\eta C^{\mu \nu }\right)  \nonumber \\
&&-2\left( 3\frac{\delta W}{\delta \varphi }H_{\lambda }^{*}B_{\mu \nu
}^{*}+W\eta _{\mu \nu \lambda }^{*}\right) C^{\mu \nu \lambda }+\left(
\left( \frac{\delta M}{\delta \varphi }C_{\rho \lambda }^{*}+\frac{\delta
^{2}M}{\delta \varphi ^{2}}H_{\rho }^{*}H_{\lambda }^{*}\right) B^{\rho
\lambda }\right.  \nonumber \\
&&\left. +2\left( \frac{\delta M}{\delta \varphi }H_{\mu }^{*}A^{*\mu
}-M\eta ^{*}\right) \right) \varepsilon _{\alpha \beta \gamma \delta }\eta
^{\alpha \beta \gamma \delta }  \nonumber \\
&&-\frac{9}{4}\left( \frac{\delta M}{\delta \varphi }C_{\rho \lambda }^{*}+%
\frac{\delta ^{2}M}{\delta \varphi ^{2}}H_{\rho }^{*}H_{\lambda }^{*}\right)
\varepsilon _{\alpha \beta \gamma \delta }\eta ^{\rho \alpha \beta }\eta
^{\lambda \gamma \delta }.  \label{bfa35}
\end{eqnarray}
Next, we investigate the equation (\ref{bfa24}) projected on antighost
number one 
\begin{equation}
\delta a_{2}+\gamma a_{1}=\partial ^{\mu }\stackrel{(1)}{m}_{\mu },
\label{bfa36}
\end{equation}
which combined with (\ref{bfa35}) further yields 
\begin{eqnarray}
&&a_{1}=\frac{\delta W}{\delta \varphi }H_{\mu }^{*}\left( 2A_{\nu }C^{\mu
\nu }-H^{\mu }\eta \right) +W\left( 2B_{\mu \nu }^{*}C^{\mu \nu }+\varphi
^{*}\eta \right)  \nonumber \\
&&+2\left( \frac{\delta M}{\delta \varphi }H_{\rho }^{*}B^{\rho \alpha
}-MA^{*\alpha }\right) \varepsilon _{\alpha \beta \gamma \delta }\eta
^{\beta \gamma \delta }+\bar{a}_{1},  \label{bfa37}
\end{eqnarray}
where 
\begin{eqnarray}
&&\bar{a}_{1}=\left( B_{\mu \nu }^{*}T^{\mu \nu }\left( \left[ \omega
^{\Delta }\right] \right) +A_{\mu }^{*}\tilde{T}^{\mu }\left( \left[ \omega
^{\Delta }\right] \right) +\varphi ^{*}T\left( \left[ \omega ^{\Delta
}\right] \right) \right.  \nonumber \\
&&\left. +H_{\mu }^{*}T^{\mu }\left( \left[ \omega ^{\Delta }\right] \right)
+y_{i}^{*}\bar{T}^{i}\left( \left[ \omega ^{\Delta }\right] \right) \right)
\eta ,  \label{bfa37a}
\end{eqnarray}
with 
\begin{equation}
\omega ^{\Delta }=\left( y^{i},\varphi ,F^{\mu \nu },\partial _{\mu }H^{\mu
},\partial _{\mu }B^{\mu \nu }\right) .  \label{bfa37b}
\end{equation}
The term (\ref{bfa37a}) added in the right hand-side of (\ref{bfa37})
appears as the general solution to the `homogeneous' equation $\gamma \bar{a}%
_{1}=0$\footnote{%
The triviality of the invariant cohomology of the exterior spacetime
differential $d$ in form degree less than four and in strictly positive
antighost number guarantees that one can always replace the equation $\gamma 
\bar{a}_{1}=\partial _{\mu }t^{\mu }$ with that corresponding to $t^{\mu }=0$%
.} and takes into account the fact that both the BF and matter theories are
involved with the local homology of the Koszul-Tate differential at
antighost number one. Its form is given by the general solution (\ref{bfa27}%
) for $I=1$. Such terms correspond to $\bar{a}_{2}=0$ and thus they do not
modify either the gauge algebra or the reducibility functions, but only the
gauge transformations of the interacting theory.

In order to solve the equation (\ref{bfa24}) at antighost number zero 
\begin{equation}
\delta a_{1}+\gamma a_{0}=\partial ^{\mu }\stackrel{(0)}{m}_{\mu },
\label{bfagh0}
\end{equation}
whose solution is nothing but the deformed Lagrangian at order one in $g$,
from (\ref{bfa37}) we observe that 
\begin{eqnarray}
&&\delta a_{1}=\partial _{\mu }\left( W\left( 2A_{\nu }C^{\mu \nu }-H^{\mu
}\eta \right) +2MB^{\mu \alpha }\varepsilon _{\alpha \beta \gamma \delta
}\eta ^{\beta \gamma \delta }\right)  \nonumber \\
&&+\gamma \left( WA_{\mu }H^{\mu }-\frac{1}{2}M\varepsilon _{\alpha \beta
\gamma \delta }B^{\alpha \beta }B^{\gamma \delta }\right) +\delta \bar{a}%
_{1}.  \label{bfa38}
\end{eqnarray}
Thus, the consistency of the deformation procedure at order one in the
coupling constant requires that $\delta \bar{a}_{1}$ must separately be $%
\gamma $-exact modulo $d$%
\begin{equation}
\delta \bar{a}_{1}+\gamma \bar{a}_{0}=\partial _{\mu }\tilde{j}^{\mu }.
\label{bfabar0}
\end{equation}
Next, we determine the concrete form of the functions $\left( T^{\mu \nu },%
\tilde{T}^{\mu },T,T^{\mu },\bar{T}^{i}\right) $ in (\ref{bfa37a}) such that
(\ref{bfabar0}) is obeyed. Recalling the definitions (\ref{bfa15}--\ref
{bfa19}), it follows that 
\begin{eqnarray}
&&\delta \bar{a}_{1}=\left( \frac{1}{2}F_{\mu \nu }T^{\mu \nu }\left( \left[
\omega ^{\Delta }\right] \right) -\left( \partial ^{\nu }B_{\nu \mu }\right) 
\tilde{T}^{\mu }\left( \left[ \omega ^{\Delta }\right] \right) -\left(
\partial _{\mu }H^{\mu }\right) T\left( \left[ \omega ^{\Delta }\right]
\right) \right.  \nonumber \\
&&\left. +\left( \partial _{\mu }\varphi \right) T^{\mu }\left( \left[
\omega ^{\Delta }\right] \right) +\left( -\right) ^{\epsilon _{i}}\frac{%
\delta ^{L}\mathcal{L}_{0}^{\mathrm{matt}}}{\delta y^{i}}\bar{T}^{i}\left(
\left[ \omega ^{\Delta }\right] \right) \right) \eta .  \label{bfabar1}
\end{eqnarray}
Taking into account the definitions (\ref{bfa21}) and the first relation
from (\ref{bfa22}), the equation (\ref{bfabar0}) possesses solutions if and
only if 
\begin{eqnarray}
&&\frac{1}{2}F_{\mu \nu }T^{\mu \nu }\left( \left[ \omega ^{\Delta }\right]
\right) -\left( \partial ^{\nu }B_{\nu \mu }\right) \tilde{T}^{\mu }\left(
\left[ \omega ^{\Delta }\right] \right) -\left( \partial _{\mu }H^{\mu
}\right) T\left( \left[ \omega ^{\Delta }\right] \right)  \nonumber \\
&&+\left( \partial _{\mu }\varphi \right) T^{\mu }\left( \left[ \omega
^{\Delta }\right] \right) +\left( -\right) ^{\epsilon _{i}}\frac{\delta ^{L}%
\mathcal{L}_{0}^{\mathrm{matt}}}{\delta y^{i}}\bar{T}^{i}\left( \left[
\omega ^{\Delta }\right] \right) =\partial _{\mu }\tilde{j}^{\mu },
\label{bfabar2}
\end{eqnarray}
where $\tilde{j}^{\mu }$ is a bosonic $\gamma $-invariant current of both
pure ghost and antighost numbers equal to zero 
\begin{equation}
\gamma \tilde{j}^{\mu }=0,\;\mathrm{pgh}\left( \tilde{j}^{\mu }\right) =0=%
\mathrm{agh}\left( \tilde{j}^{\mu }\right) ,\;\varepsilon \left( \tilde{j}%
^{\mu }\right) =0.  \label{bfabar3}
\end{equation}
The relation (\ref{bfabar2}) can easily be written under the form 
\begin{eqnarray}
&&\partial _{\mu }\left( A_{\nu }T^{\mu \nu }-B^{\mu \nu }\tilde{T}_{\nu
}-H^{\mu }T\right) +\left( \partial _{\mu }\varphi \right) T^{\mu }+\left(
-\right) ^{\epsilon _{i}}\frac{\delta ^{L}\mathcal{L}_{0}^{\mathrm{matt}}}{%
\delta y^{i}}\bar{T}^{i}  \nonumber \\
&&-A_{\nu }\partial _{\mu }T^{\mu \nu }+\frac{1}{2}B^{\mu \nu }\partial
_{\left[ \mu \right. }\tilde{T}_{\left. \nu \right] }+H^{\mu }\partial _{\mu
}T=\partial _{\mu }\tilde{j}^{\mu },  \label{bfabar4}
\end{eqnarray}
which shows that the current $\tilde{j}^{\mu }$ is given by 
\begin{equation}
\tilde{j}^{\mu }=A_{\nu }T^{\mu \nu }-B^{\mu \nu }\tilde{T}_{\nu }-H^{\mu }T+%
\bar{j}^{\mu },  \label{bfabar5}
\end{equation}
where $\bar{j}^{\mu }$ comes from $\left( \partial _{\mu }\varphi \right)
T^{\mu }+\left( -\right) ^{\epsilon _{i}}\frac{\delta ^{L}\mathcal{L}_{0}^{%
\mathrm{matt}}}{\delta y^{i}}\bar{T}^{i}$ and the functions $T^{\mu \nu }$, $%
\tilde{T}_{\nu }$ and $T$ must fulfill the condition 
\begin{equation}
-A_{\nu }\partial _{\mu }T^{\mu \nu }+\frac{1}{2}B^{\mu \nu }\partial
_{\left[ \mu \right. }\tilde{T}_{\left. \nu \right] }+H^{\mu }\partial _{\mu
}T=0.  \label{bfabar6}
\end{equation}
On behalf of the formula (\ref{bfabar5}) and using (\ref{bfa21}--\ref{bfa22}%
), we get that the equation (\ref{bfabar3}) is equivalent to 
\begin{equation}
\left( \partial _{\nu }\eta \right) T^{\mu \nu }+3\left( \partial _{\rho
}\eta ^{\mu \nu \rho }\right) \tilde{T}_{\nu }-2\left( \partial _{\nu
}C^{\mu \nu }\right) T=0,  \label{bfabar7}
\end{equation}
since all the coefficients $T^{\mu \nu }$, etc. are by assumption $\gamma $%
-closed and $\gamma \bar{j}^{\mu }=0$ by construction. Analyzing the
equations (\ref{bfabar6}--\ref{bfabar7}), we find that their solutions are
strongly equal to zero 
\begin{equation}
T^{\mu \nu }=0,\;\tilde{T}_{\nu }=0,\;T=0.  \label{bfabar8}
\end{equation}
Substituting the partial solutions (\ref{bfabar8}) back into the equation (%
\ref{bfabar2}) and also using (\ref{bfabar5}), we find that (\ref{bfabar2})
becomes 
\begin{equation}
\left( \partial _{\mu }\varphi \right) T^{\mu }\left( \left[ \omega ^{\Delta
}\right] \right) +\left( -\right) ^{\epsilon _{i}}\frac{\delta ^{L}\mathcal{L%
}_{0}^{\mathrm{matt}}}{\delta y^{i}}\bar{T}^{i}\left( \left[ \omega ^{\Delta
}\right] \right) =\partial _{\mu }\bar{j}^{\mu }.  \label{bfa38a}
\end{equation}
In order to have solutions to (\ref{bfa38a}), it is necessary to suppose
that the Lagrangian action of the matter fields is invariant under a global
one-parameter symmetry 
\begin{equation}
\Delta y^{i}=T^{i}\left( \left[ y^{j}\right] \right) \xi ,  \label{bfa40}
\end{equation}
with $\varepsilon \left( T^{i}\left( \left[ y^{j}\right] \right) \right)
=\varepsilon _{i}$ and $\xi $ a bosonic and constant parameter, which
further yields, via Noether's theorem 
\begin{equation}
\left( -\right) ^{\varepsilon _{i}}\frac{\delta ^{L}\mathcal{L}_{0}^{\mathrm{%
matt}}}{\delta y^{i}}T^{i}\left( \left[ y^{j}\right] \right) =\partial _{\mu
}j^{\mu }\left( \left[ y^{j}\right] \right) ,  \label{bfa39}
\end{equation}
the appearance of the on-shell conserved bosonic current $j^{\mu }\left(
\left[ y^{j}\right] \right) $ (on-shell means here on the stationary surface
of the field equations for the matter theory only). In the sequel we assume
that the matter theory indeed satisfies this demand. Under this assumption,
we try the unknown functions $\bar{T}^{i}\left( \left[ \omega ^{\Delta
}\right] \right) $ appearing in (\ref{bfa38a}) of the type 
\begin{equation}
\bar{T}^{i}\left( \left[ \omega ^{\Delta }\right] \right) =T^{i}\left(
\left[ y^{j}\right] \right) \bar{T}\left( \left[ \varphi \right] ,\left[
F^{\mu \nu }\right] ,\left[ \partial _{\mu }H^{\mu }\right] ,\left[ \partial
_{\mu }B^{\mu \nu }\right] \right) ,  \label{bfabar9}
\end{equation}
and thus (\ref{bfa38a}) turns, by means of Noether's theorem (\ref{bfa39}),
into 
\begin{equation}
\left( \partial _{\mu }\varphi \right) T^{\mu }\left( \left[ \omega ^{\Delta
}\right] \right) +\left( \partial _{\mu }j^{\mu }\left( \left[ y^{j}\right]
\right) \right) \bar{T}\left( \left[ \varphi \right] ,\left[ F^{\mu \nu
}\right] ,\left[ \partial _{\mu }H^{\mu }\right] ,\left[ \partial _{\mu
}B^{\mu \nu }\right] \right) =\partial _{\mu }\bar{j}^{\mu }.
\label{bfabar10}
\end{equation}
Looking at the particular form of the second term in the left hand-side of
the equation (\ref{bfabar10}), it is natural to search the functions $T^{\mu
}\left( \left[ \omega ^{\Delta }\right] \right) $ among the elements written
like 
\begin{equation}
T^{\mu }\left( \left[ \omega ^{\Delta }\right] \right) =j^{\mu }\left(
\left[ y^{j}\right] \right) \tilde{T}\left( \left[ \varphi \right] ,\left[
F^{\mu \nu }\right] ,\left[ \partial _{\mu }H^{\mu }\right] ,\left[ \partial
_{\mu }B^{\mu \nu }\right] \right) ,  \label{bfabar11}
\end{equation}
such that (\ref{bfabar10}) switches to 
\begin{eqnarray}
&&j^{\mu }\left( \left[ y^{j}\right] \right) \left( \partial _{\mu }\varphi
\right) \tilde{T}\left( \left[ \varphi \right] ,\left[ F^{\mu \nu }\right]
,\left[ \partial _{\mu }H^{\mu }\right] ,\left[ \partial _{\mu }B^{\mu \nu
}\right] \right) +  \nonumber \\
&&\left( \partial _{\mu }j^{\mu }\left( \left[ y^{j}\right] \right) \right) 
\bar{T}\left( \left[ \varphi \right] ,\left[ F^{\mu \nu }\right] ,\left[
\partial _{\mu }H^{\mu }\right] ,\left[ \partial _{\mu }B^{\mu \nu }\right]
\right) =\partial _{\mu }\bar{j}^{\mu }.  \label{bfabar12}
\end{eqnarray}
Then, it is easy to see that the general solution to the equation (\ref
{bfabar12}) reads as 
\begin{equation}
\tilde{T}\left( \left[ \varphi \right] ,\left[ F^{\mu \nu }\right] ,\left[
\partial _{\mu }H^{\mu }\right] ,\left[ \partial _{\mu }B^{\mu \nu }\right]
\right) =\frac{\delta U\left( \varphi \right) }{\delta \varphi },
\label{bfabar13}
\end{equation}
\begin{equation}
\bar{T}\left( \left[ \varphi \right] ,\left[ F^{\mu \nu }\right] ,\left[
\partial _{\mu }H^{\mu }\right] ,\left[ \partial _{\mu }B^{\mu \nu }\right]
\right) =U\left( \varphi \right) ,  \label{bfabar14}
\end{equation}
where $U\left( \varphi \right) $ is an arbitrary function of the
undifferentiated scalar field. Inserting the solutions (\ref{bfabar13}--\ref
{bfabar14}) in (\ref{bfabar9}) and (\ref{bfabar11}), we completely determine
the unknown coefficients $T^{\mu }$ and $\bar{T}^{i}$ like 
\begin{equation}
T^{\mu }\left( \left[ \omega ^{\Delta }\right] \right) =\frac{\delta U\left(
\varphi \right) }{\delta \varphi }j^{\mu }\left( \left[ y^{j}\right] \right)
,\;\bar{T}^{i}\left( \left[ \omega ^{\Delta }\right] \right) =U\left(
\varphi \right) T^{i}\left( \left[ y^{j}\right] \right) ,  \label{bfa39a}
\end{equation}
while the same solutions substituted in (\ref{bfabar12}) provide the
corresponding current $\bar{j}^{\mu }$ as 
\begin{equation}
\bar{j}^{\mu }=U\left( \varphi \right) j^{\mu }\left( \left[ y^{j}\right]
\right) .  \label{bfa39b}
\end{equation}
Replacing the expressions (\ref{bfabar8}) and (\ref{bfa39a}) in (\ref{bfa37a}%
), we obtain that the general solution to the `homogeneous' equation $\gamma 
\bar{a}_{1}=0$ that produces a consistent antighost number zero element $%
\bar{a}_{0}$ via the equation (\ref{bfabar0}) is 
\begin{equation}
\bar{a}_{1}=\left( H_{\mu }^{*}\frac{\delta U\left( \varphi \right) }{\delta
\varphi }j^{\mu }\left( \left[ y^{j}\right] \right) +y_{i}^{*}U\left(
\varphi \right) T^{i}\left( \left[ y^{j}\right] \right) \right) \eta .
\label{bfa39c}
\end{equation}
Using now (\ref{bfa39c}) and (\ref{bfa40}--\ref{bfa39}), it results that 
\begin{equation}
\delta \bar{a}_{1}=-\gamma \left( A_{\mu }U\left( \varphi \right) j^{\mu
}\left( \left[ y^{j}\right] \right) \right) +\partial _{\mu }\left( \bar{j}%
^{\mu }\eta \right) ,  \label{bfa39d}
\end{equation}
with $\bar{j}^{\mu }$ given by (\ref{bfa39b}). Introducing (\ref{bfa39d}) in
(\ref{bfa38}), we finally find that the antighost number zero component from
the first-order deformation as solution to the equation (\ref{bfagh0}) is
expressed by 
\begin{equation}
a_{0}=A_{\mu }\left( U\left( \varphi \right) j^{\mu }\left( \left[
y^{j}\right] \right) -W\left( \varphi \right) H^{\mu }\right) +\frac{1}{2}%
M\varepsilon _{\alpha \beta \gamma \delta }B^{\alpha \beta }B^{\gamma \delta
}.  \label{bfa43}
\end{equation}
In conclusion, the consistency (at antighost number zero) of the term that
mixes the matter fields and the BF ones in the first-order deformation
requires that the matter theory must be invariant under a rigid
one-parameter transformation of the type (\ref{bfa40}), which produces
Noether's theorem (\ref{bfa39}).

Strictly speaking, in (\ref{bfa43}) there also appear contributions from
some conserved currents associated with the BF field sector. Indeed, we
observe that (\ref{bfa37}) can be decomposed as 
\begin{equation}
a_{1}=a_{1}^{\mathrm{(BF)}}+a_{1}^{\mathrm{(BF-m)}},  \label{rigid1}
\end{equation}
where 
\begin{eqnarray}
&&a_{1}^{\mathrm{(BF)}}=\left( W\varphi ^{*}-\frac{\delta W}{\delta \varphi }%
H_{\mu }^{*}H^{\mu }\right) \eta +2\left( WB_{\mu \nu }^{*}+\frac{\delta W}{%
\delta \varphi }H_{\mu }^{*}A_{\nu }\right) C^{\mu \nu }  \nonumber \\
&&+2\left( \frac{\delta M}{\delta \varphi }H_{\rho }^{*}B^{\rho \alpha
}-MA^{*\alpha }\right) \varepsilon _{\alpha \beta \gamma \delta }\eta
^{\beta \gamma \delta },  \label{rigid2}
\end{eqnarray}
and 
\begin{equation}
a_{1}^{\mathrm{(BF-m)}}\equiv \bar{a}_{1}=\left( H_{\mu }^{*}\frac{\delta
U\left( \varphi \right) }{\delta \varphi }j^{\mu }\left( \left[ y^{j}\right]
\right) +y_{i}^{*}U\left( \varphi \right) T^{i}\left( \left[ y^{j}\right]
\right) \right) \eta .  \label{rigid2a}
\end{equation}
We have seen in the above that $\delta a_{1}^{\mathrm{(BF-m)}}$ leads to the
term $A_{\mu }Uj^{\mu }$ from $a_{0}$, while $\delta a_{1}^{\mathrm{(BF)}}$
produces the rest of the terms in $a_{0}$, such that $a_{1}^{\mathrm{(BF-m)}%
} $ and $a_{1}^{\mathrm{(BF)}}$ are separately consistent. According to the
definitions (\ref{bfa15}--\ref{bfa19}), the Koszul-Tate differential acting
on $a_{1}^{\mathrm{(BF)}}$ gives 
\begin{eqnarray}
&&\delta a_{1}^{\mathrm{(BF)}}=\left( \frac{\delta \mathcal{L}_{0}^{\mathrm{%
BF}}}{\delta \varphi }W-\frac{\delta \mathcal{L}_{0}^{\mathrm{BF}}}{\delta
H^{\mu }}\frac{\delta W}{\delta \varphi }H^{\mu }\right) \eta +\left( 2\frac{%
\delta \mathcal{L}_{0}^{\mathrm{BF}}}{\delta B^{\mu \nu }}W+\frac{\delta 
\mathcal{L}_{0}^{\mathrm{BF}}}{\delta H^{\left[ \mu \right. }}A_{\left. \nu
\right] }\frac{\delta W}{\delta \varphi }\right) C^{\mu \nu }  \nonumber \\
&&+2\left( \frac{\delta \mathcal{L}_{0}^{\mathrm{BF}}}{\delta H^{\rho }}%
\frac{\delta M}{\delta \varphi }B^{\rho \alpha }-\frac{\delta \mathcal{L}%
_{0}^{\mathrm{BF}}}{\delta A_{\alpha }}M\right) \varepsilon _{\alpha \beta
\gamma \delta }\eta ^{\beta \gamma \delta },  \label{rigid3}
\end{eqnarray}
where $\mathcal{L}_{0}^{\mathrm{BF}}$ is given in (\ref{bfa1}) 
\begin{equation}
\mathcal{L}_{0}^{\mathrm{BF}}=H_{\mu }\partial ^{\mu }\varphi +\frac{1}{2}%
B^{\mu \nu }\partial _{[\mu }A_{\nu ]}.  \label{rigid3a}
\end{equation}
The consistency of the first-order deformation for the BF sector at
antighost number zero, $\delta a_{1}^{\mathrm{(BF)}}+\gamma a_{0}^{\mathrm{%
(BF)}}=\partial _{\rho }m^{\mathrm{(BF)}\rho }$, requires that 
\begin{equation}
\frac{\delta \mathcal{L}_{0}^{\mathrm{BF}}}{\delta \varphi }W-\frac{\delta 
\mathcal{L}_{0}^{\mathrm{BF}}}{\delta H^{\mu }}\frac{\delta W}{\delta
\varphi }H^{\mu }=\partial _{\rho }k^{\rho },  \label{rigid4}
\end{equation}
\begin{equation}
\left( 2\frac{\delta \mathcal{L}_{0}^{\mathrm{BF}}}{\delta B^{\mu \nu }}W+%
\frac{\delta \mathcal{L}_{0}^{\mathrm{BF}}}{\delta H^{\left[ \mu \right. }}%
A_{\left. \nu \right] }\frac{\delta W}{\delta \varphi }\right) =\partial
_{\rho }k_{\;\;\mu \nu }^{\rho },  \label{rigid5}
\end{equation}
\begin{equation}
2\left( \frac{\delta \mathcal{L}_{0}^{\mathrm{BF}}}{\delta H^{\rho }}\frac{%
\delta M}{\delta \varphi }B^{\rho \alpha }-\frac{\delta \mathcal{L}_{0}^{%
\mathrm{BF}}}{\delta A_{\alpha }}M\right) \varepsilon _{\alpha \beta \gamma
\delta }=\partial _{\rho }k_{\;\;\beta \gamma \delta }^{\rho },
\label{rigid6}
\end{equation}
where, in addition, we must have $k_{\;\;\mu \nu }^{\rho }=\delta
_{\;\;\left[ \mu \right. }^{\rho }\bar{k}_{\left. \nu \right] }$ and $%
k_{\;\;\beta \gamma \delta }^{\rho }=\delta _{\;\;\left[ \beta \right.
}^{\rho }\bar{k}_{\left. \gamma \delta \right] }$, for some $\bar{k}_{\nu }$
and antisymmetric $\bar{k}_{\gamma \delta }$. After some computation, we
infer that (\ref{rigid4}--\ref{rigid6}) hold, and they are nothing but
Noether's theorem expressing the invariance of the pure BF theory
respectively under three different non-trivial rigid symmetries 
\begin{equation}
\Delta _{1}\varphi =W\left( \varphi \right) \xi ^{\prime },\;\Delta
_{1}H^{\mu }=-\frac{\delta W}{\delta \varphi }H^{\mu }\xi ^{\prime
},\;\Delta _{1}A^{\mu }=\Delta _{1}B^{\mu \nu }=0,  \label{rigid7}
\end{equation}
\begin{equation}
\Delta _{2}B^{\mu \nu }=W\left( \varphi \right) \delta _{\left[ \alpha
\right. }^{\mu }\delta _{\left. \beta \right] }^{\nu }\xi ^{\alpha \beta
},\;\Delta _{2}H^{\mu }=\frac{\delta W}{\delta \varphi }\delta _{\left[
\alpha \right. }^{\mu }A_{\left. \beta \right] }\xi ^{\alpha \beta
},\;\Delta _{2}A^{\mu }=\Delta _{2}\varphi =0,  \label{rigid8}
\end{equation}
\begin{equation}
\Delta _{3}H^{\rho }=2\frac{\delta M}{\delta \varphi }B^{\rho \alpha
}\varepsilon _{\alpha \beta \gamma \delta }\xi ^{\beta \gamma \delta
},\;\Delta _{3}A_{\alpha }=-2M\left( \varphi \right) \varepsilon _{\alpha
\beta \gamma \delta }\xi ^{\beta \gamma \delta },  \label{oricr}
\end{equation}
\begin{equation}
\Delta _{3}B^{\mu \nu }=\Delta _{3}\varphi =0,  \label{rigid9}
\end{equation}
that result in the conserved currents 
\begin{equation}
k^{\rho }=-W\left( \varphi \right) H^{\rho },  \label{rigid10}
\end{equation}
\begin{equation}
k_{\;\;\alpha \beta }^{\rho }=W\left( \varphi \right) \delta _{\left[ \alpha
\right. }^{\rho }A_{\left. \beta \right] },  \label{rigid11}
\end{equation}
respectively 
\begin{equation}
k_{\;\;\beta \gamma \delta }^{\rho }=2M\left( \varphi \right) B^{\rho \alpha
}\varepsilon _{\alpha \beta \gamma \delta }.  \label{rigid12}
\end{equation}
As a consequence, we infer that the deformed Lagrangian at order one in the
coupling constant, (\ref{bfa43}), splits as 
\begin{equation}
a_{0}=a_{0}^{\mathrm{(BF)}}+a_{0}^{\mathrm{(BF-m)}},  \label{rigid13}
\end{equation}
with 
\begin{equation}
a_{0}^{\mathrm{(BF)}}=-W\left( \varphi \right) A_{\mu }H^{\mu }+\frac{1}{2}%
M\left( \varphi \right) \varepsilon _{\alpha \beta \gamma \delta }B^{\alpha
\beta }B^{\gamma \delta },  \label{rigid14}
\end{equation}
\begin{equation}
a_{0}^{\mathrm{(BF-m)}}=U\left( \varphi \right) A_{\mu }j^{\mu }\left(
\left[ y^{j}\right] \right) ,  \label{rigid14a}
\end{equation}
where $a_{0}^{\mathrm{(BF)}}$ can equivalently be written in terms of the
currents (\ref{rigid10}--\ref{rigid12}) under the form 
\begin{equation}
a_{0}^{\mathrm{(BF)}}=pk^{\rho }A_{\rho }-\frac{1-p}{3}k_{\;\;\rho \beta
}^{\rho }H^{\beta }-\frac{1}{4}k_{\;\;\rho \gamma \delta }^{\rho }B^{\gamma
\delta },  \label{rigid}
\end{equation}
where $p$ is a real number. Nevertheless, these non-trivial currents appear
in a natural way within the cohomological framework used here, and do not
need to be assumed to exist, unlike the matter fields, whose non-trivial
couplings to the BF field sector require that the matter theory indeed
possesses non-trivial rigid one-parameter symmetries.

Let us briefly discuss the link between the set of rigid symmetries and the
local homology of the Koszul-Tate differential. The equations (\ref{bfa39})
and (\ref{rigid4}--\ref{rigid6}) may be rewritten in terms of the
Koszul-Tate differential as 
\begin{equation}
\partial _{\mu }j_{\;\;\Delta }^{\mu }=\delta \left( -\Phi _{\alpha
_{0}}^{*}T_{\;\;\Delta }^{\alpha _{0}}\right) \equiv \delta \sigma _{\Delta
},  \label{bfa41}
\end{equation}
where $j_{\;\;\Delta }^{\mu }$ and $T_{\;\;\Delta }^{\alpha _{0}}$
collectively denote the corresponding conserved currents, respectively, the
generators of the rigid symmetries. The formula (\ref{bfa41}) correlates the
rigid symmetries (\ref{bfa40}) and (\ref{rigid7}--\ref{rigid9}) to certain
homological classes from the space $H_{1}\left( \delta |d\right) $.
Explicitly, it shows that a global symmetry (materialized in a conserved
current) defines an element $\sigma _{\Delta }$ of $H_{1}\left( \delta
|d\right) $, i.e., an element of antighost number equal to one that is $%
\delta $-closed modulo $d$. A global symmetry is said to be trivial if the
corresponding $\sigma _{\Delta }$ is in a trivial class of $H_{1}\left(
\delta |d\right) $, hence if it is $\delta $-exact modulo $d$%
\begin{equation}
\sigma _{\Delta }=\delta \rho _{\Delta }+\partial _{\mu }c_{\Delta }^{\mu
},\;\mathrm{antigh}\left( \rho _{\Delta }\right) =2,\;\mathrm{antigh}\left(
c_{\Delta }^{\mu }\right) =1.  \label{bfa42}
\end{equation}
A class of trivial rigid symmetries corresponds to some rigid generators of
the type 
\begin{equation}
T_{\;\;\Delta }^{\alpha _{0}}=M_{\Delta }^{\alpha _{0}\beta _{0}}\left(
\left[ \Phi ^{\gamma _{0}}\right] \right) \frac{\delta ^{L}\mathcal{L}}{%
\delta \Phi ^{\beta _{0}}},\;M_{\Delta }^{\alpha _{0}\beta _{0}}=-\left(
-\right) ^{\varepsilon _{\alpha _{0}}\varepsilon _{\beta _{0}}}M_{\Delta
}^{\beta _{0}\alpha _{0}},  \label{rigidz}
\end{equation}
where $\mathcal{L}=\mathcal{L}_{0}^{\mathrm{BF}}+\mathcal{L}_{0}^{\mathrm{%
matt}}\left( \left[ y^{i}\right] \right) $. In (\ref{rigidz}) $\varepsilon
_{\alpha _{0}}$ is the Grassmann parity of the field $\Phi ^{\alpha _{0}}$,
such that $\varepsilon \left( M_{\Delta }^{\alpha _{0}\beta _{0}}\right)
=\left( \varepsilon _{\alpha _{0}}+\varepsilon _{\beta _{0}}\right) \mathrm{%
mod}\;2$ as all the rigid parameters are assumed to be bosonic. In this
situation we have that 
\begin{equation}
\sigma _{\Delta }=-\Phi _{\alpha _{0}}^{*}T_{\;\;\Delta }^{\alpha
_{0}}=\delta \left( \frac{1}{2}\Phi _{\alpha _{0}}^{*}M_{\Delta }^{\alpha
_{0}\beta _{0}}\Phi _{\beta _{0}}^{*}\right) ,  \label{rigidy}
\end{equation}
and thus $\sigma _{\Delta }$ is trivial in $H_{1}\left( \delta |d\right) $.
Noether's theorem for trivial rigid symmetries 
\begin{equation}
\left( -\right) ^{\varepsilon _{\alpha _{0}}}\frac{\delta ^{L}\mathcal{L}}{%
\delta \Phi ^{\alpha _{0}}}T_{\;\;\Delta }^{\alpha _{0}}=\partial _{\mu
}j_{\;\;\Delta }^{\mu },  \label{rigidw}
\end{equation}
obviously reduces to $\partial _{\mu }j_{\;\;\Delta }^{\mu }=0$, which holds
independently of the field equations. Then, we further obtain that 
\begin{equation}
j_{\;\;\Delta }^{\mu }=\partial _{\nu }t_{\;\;\Delta }^{\mu \nu
},\;t_{\;\;\Delta }^{\mu \nu }=-t_{\;\;\Delta }^{\nu \mu },  \label{rigidl}
\end{equation}
such that the currents associated with trivial global symmetries are also
trivial. We remark that the rigid symmetries (\ref{rigid7}--\ref{rigid9})
are not of the type (\ref{rigidz}), and therefore are non-trivial.

In order to effectively couple the matter fields to the BF ones, we suppose
that the matter rigid symmetries are also non-trivial. Indeed, if the matter
rigid symmetries were trivial 
\begin{equation}
T^{i}=M^{ij}\left( \left[ y^{k}\right] \right) \frac{\delta ^{L}\mathcal{L}%
_{0}^{\mathrm{matt}}}{\delta y^{j}},\;M^{ij}=-\left( -\right) ^{\varepsilon
_{i}\varepsilon _{j}}M^{ji},  \label{rigidm}
\end{equation}
then the conserved current associated with (\ref{rigidm}) is also trivial, 
\begin{equation}
j^{\mu }=\partial _{\nu }t^{\mu \nu },\;t^{\mu \nu }=-t^{\nu \mu },
\label{trivy}
\end{equation}
such that the terms $a_{1}^{\mathrm{(BF-m)}}+a_{0}^{\mathrm{(BF-m)}}$ from
the first-order deformation are $s$-exact modulo $d$%
\begin{eqnarray}
&&a_{1}^{\mathrm{(BF-m)}}+a_{0}^{\mathrm{(BF-m)}}\rightarrow \left( H_{\mu
}^{*}\frac{\delta U\left( \varphi \right) }{\delta \varphi }\eta +U\left(
\varphi \right) A_{\mu }\right) \partial _{\nu }t^{\mu \nu }\left( \left[
y^{j}\right] \right)  \nonumber \\
&&+y_{i}^{*}U\left( \varphi \right) M^{ij}\left( \left[ y^{k}\right] \right) 
\frac{\delta ^{L}\mathcal{L}_{0}^{\mathrm{matt}}}{\delta y^{j}}\eta =s\left(
-t^{\mu \nu }\left( \left[ y^{j}\right] \right) \left( \frac{1}{2}\frac{%
\delta ^{2}U\left( \varphi \right) }{\delta \varphi ^{2}}H_{\mu }^{*}H_{\nu
}^{*}\eta \right. \right.  \nonumber \\
&&\left. \left. +\frac{\delta U\left( \varphi \right) }{\delta \varphi }%
\left( \frac{1}{2}C_{\mu \nu }^{*}\eta +H_{\mu }^{*}A_{\nu }\right) +B_{\mu
\nu }^{*}U\left( \varphi \right) \right) +\frac{1}{2}U\left( \varphi \right)
y_{i}^{*}M^{ij}\left( \left[ y^{k}\right] \right) y_{j}^{*}\eta \right) 
\nonumber \\
&&+\partial _{\nu }\left( t^{\mu \nu }\left( \left[ y^{j}\right] \right)
\left( \frac{\delta U\left( \varphi \right) }{\delta \varphi }H_{\mu
}^{*}\eta +U\left( \varphi \right) A_{\mu }\right) \right) .  \label{rigidkn}
\end{eqnarray}
Since the first-order deformation is unique up to $s$-exact terms plus
divergences of local currents, in the case of trivial matter rigid
symmetries we can remove the pieces $a_{1}^{\mathrm{(BF-m)}}+a_{0}^{\mathrm{%
(BF-m)}}$ from $a$, and thus in this situation there are no couplings
between the matter and the BF field sectors.

Combining the formulas (\ref{bfa31}--\ref{bfa33}) and (\ref{bfa35}) with the
expression (\ref{bfa37}) in which we use the solution (\ref{bfa39c}), and
also with the result given by (\ref{bfa43}), we conclude that the
first-order deformation of the solution to the master equation for the model
under study can be written in the form 
\begin{eqnarray}
&&S_{1}=\int d^{4}x\left( A_{\mu }\left( U\left( \varphi \right) j^{\mu
}\left( \left[ y^{j}\right] \right) -W\left( \varphi \right) H^{\mu }\right)
+\frac{1}{2}M\varepsilon _{\alpha \beta \gamma \delta }B^{\alpha \beta
}B^{\gamma \delta }\right.  \nonumber \\
&&+H_{\mu }^{*}\left( 2\frac{\delta W}{\delta \varphi }A_{\nu }C^{\mu \nu
}+\left( \frac{\delta U}{\delta \varphi }j^{\mu }\left( \left[ y^{j}\right]
\right) -\frac{\delta W}{\delta \varphi }H^{\mu }\right) \eta \right) 
\nonumber \\
&&+W\left( \varphi \right) \left( 2B_{\mu \nu }^{*}C^{\mu \nu }+\varphi
^{*}\eta \right) +y_{i}^{*}U\left( \varphi \right) T^{i}\left( \left[
y^{j}\right] \right) \eta  \nonumber \\
&&+2\left( \frac{\delta M}{\delta \varphi }H_{\rho }^{*}B^{\rho \alpha
}-M\left( \varphi \right) A^{*\alpha }\right) \varepsilon _{\alpha \beta
\gamma \delta }\eta ^{\beta \gamma \delta }  \nonumber \\
&&+\left( \frac{\delta W}{\delta \varphi }C_{\mu \nu }^{*}+\frac{\delta ^{2}W%
}{\delta \varphi ^{2}}H_{\mu }^{*}H_{\nu }^{*}\right) \left( -3A_{\lambda
}C^{\mu \nu \lambda }+\eta C^{\mu \nu }\right)  \nonumber \\
&&-2\left( 3\frac{\delta W}{\delta \varphi }H_{\lambda }^{*}B_{\mu \nu
}^{*}+W\left( \varphi \right) \eta _{\mu \nu \lambda }^{*}\right) C^{\mu \nu
\lambda }+\left( \left( \frac{\delta M}{\delta \varphi }C_{\rho \lambda
}^{*}\right. \right.  \nonumber \\
&&\left. \left. +\frac{\delta ^{2}M}{\delta \varphi ^{2}}H_{\rho
}^{*}H_{\lambda }^{*}\right) B^{\rho \lambda }+2\left( \frac{\delta M}{%
\delta \varphi }H_{\mu }^{*}A^{*\mu }-M\left( \varphi \right) \eta
^{*}\right) \right) \varepsilon _{\alpha \beta \gamma \delta }\eta ^{\alpha
\beta \gamma \delta }  \nonumber \\
&&-\frac{9}{4}\left( \frac{\delta M}{\delta \varphi }C_{\rho \lambda }^{*}+%
\frac{\delta ^{2}M}{\delta \varphi ^{2}}H_{\rho }^{*}H_{\lambda }^{*}\right)
\varepsilon _{\alpha \beta \gamma \delta }\eta ^{\rho \alpha \beta }\eta
^{\lambda \gamma \delta }  \nonumber \\
&&-\left( \frac{\delta W}{\delta \varphi }C_{\nu \rho \lambda }^{*}+\frac{%
\delta ^{2}W}{\delta \varphi ^{2}}H_{\left[ \nu \right. }^{*}C_{\left. \rho
\lambda \right] }^{*}+\frac{\delta ^{3}W}{\delta \varphi ^{3}}H_{\nu
}^{*}H_{\rho }^{*}H_{\lambda }^{*}\right) \times  \nonumber \\
&&\times \left( 4A_{\mu }C^{\mu \nu \rho \lambda }+\eta C^{\nu \rho \lambda
}\right) +2W\left( \varphi \right) \eta _{\mu \nu \rho \lambda }^{*}C^{\mu
\nu \rho \lambda }-8\frac{\delta W}{\delta \varphi }H_{\lambda }^{*}\eta
_{\mu \nu \rho }^{*}C^{\mu \nu \rho \lambda }  \nonumber \\
&&+12\left( \frac{\delta W}{\delta \varphi }C_{\rho \lambda }^{*}+\frac{%
\delta ^{2}W}{\delta \varphi ^{2}}H_{\rho }^{*}H_{\lambda }^{*}\right)
B_{\mu \nu }^{*}C^{\mu \nu \rho \lambda }-\left( \frac{\delta M}{\delta
\varphi }C_{\nu \rho \lambda }^{*}\right.  \nonumber \\
&&\left. +\frac{\delta ^{2}M}{\delta \varphi ^{2}}H_{\left[ \nu \right.
}^{*}C_{\left. \rho \lambda \right] }^{*}+\frac{\delta ^{3}M}{\delta \varphi
^{3}}H_{\nu }^{*}H_{\rho }^{*}H_{\lambda }^{*}\right) \eta ^{\nu \rho
\lambda }\varepsilon _{\alpha \beta \gamma \delta }\eta ^{\alpha \beta
\gamma \delta }  \nonumber \\
&&+\left( \frac{\delta W}{\delta \varphi }C_{\mu \nu \rho \lambda }^{*}+%
\frac{\delta ^{2}W}{\delta \varphi ^{2}}\left( H_{\left[ \mu \right.
}^{*}C_{\left. \nu \rho \lambda \right] }^{*}+C_{\left[ \mu \nu \right.
}^{*}C_{\left. \rho \lambda \right] }^{*}\right) \right.  \nonumber \\
&&\left. +\frac{\delta ^{3}W}{\delta \varphi ^{3}}H_{\left[ \mu
\right. }^{*}H_{\nu }^{*}C_{\left. \rho \lambda \right] }^{*}+\frac{\delta
^{4}W}{\delta \varphi ^{4}}H_{\mu }^{*}H_{\nu }^{*}H_{\rho }^{*}H_{\lambda
}^{*}\right) \eta C^{\mu \nu \rho \lambda } \nonumber \\
&&+\frac{1}{2}\left( \frac{\delta M}{\delta \varphi }C_{\mu \nu \rho \lambda
}^{*}+\frac{\delta ^{2}M}{\delta \varphi ^{2}}\left( H_{\left[ \mu \right.
}^{*}C_{\left. \nu \rho \lambda \right] }^{*}+C_{\left[ \mu \nu \right.
}^{*}C_{\left. \rho \lambda \right] }^{*}\right) \right.  \nonumber \\
&&\left. \left. +\frac{\delta ^{3}M}{\delta \varphi ^{3}}H_{\left[ \mu \right.
}^{*}H_{\nu }^{*}C_{\left. \rho \lambda \right] }^{*}+\frac{\delta ^{4}M}{%
\delta \varphi ^{4}}H_{\mu }^{*}H_{\nu }^{*}H_{\rho }^{*}H_{\lambda
}^{*}\right) \eta ^{\mu \nu \rho \lambda }\varepsilon _{\alpha \beta \gamma
\delta }\eta ^{\alpha \beta \gamma \delta } \right) .  \label{bfa44}
\end{eqnarray}
It is by construction a $s$-cocycle of ghost number zero, such that $%
S_{0}+gS_{1}$ is solution to the master equation to order $g$.

\subsection{Higher-order deformations}

Next, we investigate the equations that control the higher-order
deformations. The second-order deformation is governed by the equation (\ref
{bfa2.6}). Making use of (\ref{bfa44}), the second term in the left
hand-side of (\ref{bfa2.6}) takes the concrete form 
\begin{eqnarray}
&&\frac{1}{2}\left( S_{1},S_{1}\right) =\varepsilon _{\mu \nu \rho \lambda
}\sum\limits_{a=0}^{4}\int d^{4}x\left( T_{a}^{\mu \nu \rho \lambda }\frac{%
\delta ^{a}X}{\delta \varphi ^{a}}+U_{a}^{\mu \nu \rho \lambda }\frac{\delta
^{a}Y}{\delta \varphi ^{a}}\right)  \nonumber \\
&&-2\varepsilon _{\mu \nu \rho \lambda }\int d^{4}x\left( \left( \frac{%
\delta \left( MU\right) }{\delta \varphi }j^{\alpha }H_{\alpha
}^{*}+MUy_{i}^{*}T^{i}\right) \eta ^{\mu \nu \rho \lambda }+MUj^{\mu }\eta
^{\nu \rho \lambda }\right)  \nonumber \\
&&+\int d^{4}xU^{2}\frac{\delta ^{R}j^{\mu }}{\delta y^{i}}T^{i}A_{\mu }\eta
,  \label{bfa45}
\end{eqnarray}
where 
\begin{equation}
T_{0}^{\mu \nu \rho \lambda }=4A^{*\mu }C^{\nu \rho \lambda }+B^{\mu \nu
}C^{\rho \lambda }+H^{\mu }\eta ^{\nu \rho \lambda }-2\eta ^{*}C^{\mu \nu
\rho \lambda }-\varphi ^{*}\eta ^{\mu \nu \rho \lambda },  \label{ide1}
\end{equation}
\begin{eqnarray}
&&T_{1}^{\mu \nu \rho \lambda }=\left( H_{\alpha }^{*}H^{\alpha }+C_{\alpha
\beta }^{*}C^{\alpha \beta }+C_{\alpha \beta \gamma }^{*}C^{\alpha \beta
\gamma }+C_{\alpha \beta \gamma \delta }^{*}C^{\alpha \beta \gamma \delta
}\right) \eta ^{\mu \nu \rho \lambda }  \nonumber \\
&&+\left( 2H_{\alpha }^{*}A^{*\alpha }+C_{\alpha \beta }^{*}B^{\alpha \beta
}-C_{\alpha \beta \gamma }^{*}\eta ^{\alpha \beta \gamma }\right) C^{\mu \nu
\rho \lambda }  \nonumber \\
&&+\left( 3C_{\alpha \beta }^{*}C^{\alpha \beta \mu }-2H_{\alpha
}^{*}C^{\alpha \mu }\right) \eta ^{\nu \rho \lambda }+3H_{\alpha
}^{*}C^{\alpha \mu \nu }B^{\rho \lambda },  \label{ide2}
\end{eqnarray}
\begin{eqnarray}
&&T_{2}^{\mu \nu \rho \lambda }=H_{\alpha }^{*}\left( \left( H_{\beta
}^{*}B^{\alpha \beta }-3C_{\beta \gamma }^{*}\eta ^{\alpha \beta \gamma
}\right) C^{\mu \nu \rho \lambda }+3H_{\beta }^{*}C^{\alpha \beta \mu }\eta
^{\nu \rho \lambda }\right)  \nonumber \\
&&+\left( \left( 4H_{\alpha }^{*}C_{\beta \gamma \delta }^{*}+3C_{\alpha
\beta }^{*}C_{\gamma \delta }^{*}\right) C^{\alpha \beta \gamma \delta
}\right.  \nonumber \\
&&\left. +H_{\alpha }^{*}\left( 3C_{\beta \gamma }^{*}C^{\alpha \beta \gamma
}+H_{\beta }^{*}C^{\alpha \beta }\right) \right) \eta ^{\mu \nu \rho \lambda
},  \label{ide3}
\end{eqnarray}
\begin{equation}
T_{3}^{\mu \nu \rho \lambda }=H_{\alpha }^{*}H_{\beta }^{*}\left( \left(
H_{\gamma }^{*}C^{\alpha \beta \gamma }+3C_{\gamma \delta }^{*}C^{\alpha
\beta \gamma \delta }\right) \eta ^{\mu \nu \rho \lambda }-H_{\gamma
}^{*}\eta ^{\alpha \beta \gamma }C^{\mu \nu \rho \lambda }\right) ,
\label{ide4}
\end{equation}
\begin{equation}
T_{4}^{\mu \nu \rho \lambda }=H_{\alpha }^{*}H_{\beta }^{*}H_{\gamma
}^{*}H_{\delta }^{*}C^{\alpha \beta \gamma \delta }\eta ^{\mu \nu \rho
\lambda },  \label{ide5}
\end{equation}
\begin{eqnarray}
&&U_{0}^{\mu \nu \rho \lambda }=\left( \frac{1}{2}\eta _{\alpha \beta \gamma
\delta }^{*}\eta ^{\alpha \beta \gamma \delta }+\eta _{\alpha \beta \gamma
}^{*}\eta ^{\alpha \beta \gamma }+B_{\alpha \beta }^{*}B^{\alpha \beta
}-6A_{\alpha }^{*}A^{\alpha }\right) \eta ^{\mu \nu \rho \lambda }  \nonumber
\\
&&+\left( A^{*\mu }\eta +\frac{3}{2}B_{\alpha \beta }^{*}\eta ^{\alpha \beta
\mu }-A_{\alpha }B^{\alpha \mu }\right) \eta ^{\nu \rho \lambda }+\frac{1}{2}%
B^{\mu \nu }B^{\rho \lambda }\eta ,  \label{ide6}
\end{eqnarray}
\begin{eqnarray}
&&U_{1}^{\mu \nu \rho \lambda }=\left( \frac{1}{4}\eta C^{*\mu \nu \rho
\lambda }-A^{\mu }C^{*\nu \rho \lambda }+3B^{*\mu \nu }C^{*\rho \lambda
}-2\eta ^{*\mu \nu \rho }H^{*\lambda }\right) \eta _{\alpha \beta \gamma
\delta }\eta ^{\alpha \beta \gamma \delta }  \nonumber \\
&&+\left( \left( \frac{1}{2}C_{\alpha \beta \gamma }^{*}\eta +\frac{3}{2}%
C_{\alpha \beta }^{*}A_{\gamma }-3B_{\alpha \beta }^{*}H_{\gamma
}^{*}\right) \eta ^{\alpha \beta \gamma }+\left( \frac{1}{2}C_{\alpha \beta
}^{*}B^{\alpha \beta }+A^{*\alpha }H_{\alpha }^{*}\right) \eta \right. 
\nonumber \\
&&\left. +H_{\alpha }^{*}A_{\beta }B^{\alpha \beta }\right) \eta ^{\mu \nu
\rho \lambda }+\left( \frac{3}{2}\left( \frac{1}{2}C_{\alpha \beta }^{*}\eta
+H_{\alpha }^{*}A_{\beta }\right) \eta ^{\alpha \beta \mu }-H_{\alpha
}^{*}B^{\alpha \mu }\eta \right) \eta ^{\nu \rho \lambda },  \label{ide7}
\end{eqnarray}
\begin{eqnarray}
&&U_{2}^{\mu \nu \rho \lambda }=H_{\alpha }^{*}\left( \frac{3}{2}\left(
C_{\beta \gamma }^{*}\eta +H_{\beta }^{*}A_{\gamma }\right) \eta ^{\alpha
\beta \gamma }+H_{\beta }^{*}B^{\alpha \beta }\eta \right) \eta ^{\mu \nu
\rho \lambda }  \nonumber \\
&&+\frac{3}{4}H_{\alpha }^{*}H_{\beta }^{*}\eta \eta ^{\alpha \beta \mu
}\eta ^{\nu \lambda \rho }+\left( H^{*\mu }\left( C^{*\nu \rho \lambda }\eta
+3H^{*\nu }B^{*\rho \lambda }\right) \right.  \nonumber \\
&&\left. +3\left( \frac{1}{4}C^{*\mu \nu }\eta +H^{*\mu }A^{\nu }\right)
C^{*\rho \lambda }\right) \eta _{\alpha \beta \gamma \delta }\eta ^{\alpha
\beta \gamma \delta },  \label{ide8}
\end{eqnarray}
\begin{eqnarray}
&&U_{3}^{\mu \nu \rho \lambda }=\frac{1}{2}H^{*\mu }H^{*\nu }\left(
3C^{*\rho \lambda }\eta +2H^{*\rho }A^{\lambda }\right) \eta _{\alpha \beta
\gamma \delta }\eta ^{\alpha \beta \gamma \delta }  \nonumber \\
&&+\frac{1}{2}H_{\alpha }^{*}H_{\beta }^{*}H_{\gamma }^{*}\eta \eta ^{\alpha
\beta \gamma }\eta ^{\mu \nu \rho \lambda },  \label{ide9}
\end{eqnarray}
\begin{equation}
U_{4}^{\mu \nu \rho \lambda }=\frac{1}{2}\eta H^{*\mu }H^{*\nu }H^{*\rho
}H^{*\lambda }\eta _{\alpha \beta \gamma \delta }\eta ^{\alpha \beta \gamma
\delta },  \label{ide10}
\end{equation}
and 
\begin{equation}
X\left( \varphi \right) =W\left( \varphi \right) M\left( \varphi \right)
,\;Y\left( \varphi \right) =W\left( \varphi \right) \frac{\delta M\left(
\varphi \right) }{\delta \varphi }.  \label{ide11}
\end{equation}
It is clear that none of the terms involving any of the functions $X$, $Y$
or their derivatives with respect to the scalar field can be written as
required by the equation (\ref{bfa2.6}), namely, like the $s$-variation of
some local functional, and therefore they must vanish. This takes place if
and only if 
\begin{equation}
W\left( \varphi \right) M\left( \varphi \right) =0.  \label{ide12}
\end{equation}
Thus, there appear two alternatives.

(I) First, we assume that 
\begin{equation}
M\left( \varphi \right) =0,  \label{ide13}
\end{equation}
and $W\left( \varphi \right) $ is an arbitrary function of the
undifferentiated scalar field. In this situation, we find that (\ref{bfa45})
reduces to 
\begin{equation}
\frac{1}{2}\left( S_{1},S_{1}\right) =\int d^{4}x\left( U\left( \varphi
\right) \right) ^{2}\frac{\delta ^{R}j^{\mu }}{\delta y^{i}}T^{i}\left(
\left[ y^{j}\right] \right) A_{\mu }\eta ,  \label{ide14}
\end{equation}
where we assume that the arbitrary function $U$ of the undifferentiated
scalar field, to be called the `background' potential in what follows, is
non-vanishing. Two major situations met in practical applications deserve
special attention.

(I.1) It might happen that the matter current is invariant under the gauge
version of the global one-parameter symmetry (\ref{bfa40}) 
\begin{equation}
\frac{\delta ^{R}j^{\mu }\left( \left[ y^{j}\right] \right) }{\delta y^{i}}%
T^{i}\left( \left[ y^{k}\right] \right) =0.  \label{bfa46}
\end{equation}
In this case it results that $\left( S_{1},S_{1}\right) =0$, such that we
can set $S_{2}=0$. Further, all the higher-order equations (\ref{bfa2.7}),
etc., are satisfied with the choice $S_{3}=S_{4}=\cdots =0$. Consequently,
the deformed solution to the master equation consistent to all orders in the
coupling constant reduces in this situation to the sum between the ``free''
solution (\ref{bfa14}) and the first-order deformation (\ref{bfa44}) where
we set $M\left( \varphi \right) =0$%
\begin{eqnarray}
&&S=\int d^{4}x\left( H^{\mu }\left( \partial _{\mu }\varphi -gW\left(
\varphi \right) A_{\mu }\right) +\frac{1}{2}B^{\mu \nu }\partial _{[\mu
}A_{\nu ]}+gU\left( \varphi \right) j^{\mu }A_{\mu }\right.  \nonumber \\
&&+\mathcal{L}_{0}^{\mathrm{matt}}\left( y^{i},\partial _{\mu }y^{i},\cdots
\right) +gy_{i}^{*}U\left( \varphi \right) T^{i}\left( \left[ y^{j}\right]
\right) \eta +A^{*\mu }\partial _{\mu }\eta  \nonumber \\
&&+H_{\mu }^{*}\left( 2\left( \partial _{\nu }+g\frac{\delta W}{\delta
\varphi }A_{\nu }\right) C^{\mu \nu }+g\left( \frac{\delta U}{\delta \varphi 
}j^{\mu }\left( \left[ y^{j}\right] \right) -\frac{\delta W}{\delta \varphi }%
H^{\mu }\right) \eta \right)  \nonumber \\
&&+B_{\mu \nu }^{*}\left( -3\partial _{\rho }\eta ^{\mu \nu \rho }+2gW\left(
\varphi \right) C^{\mu \nu }\right) +g\varphi ^{*}W\left( \varphi \right)
\eta  \nonumber \\
&&+4\eta _{\mu \nu \lambda }^{*}\partial _{\rho }\eta ^{\mu \nu \lambda \rho
}-\left( 3C_{\mu \nu }^{*}\left( \partial _{\lambda }+g\frac{\delta W}{%
\delta \varphi }A_{\lambda }\right) +2g\eta _{\mu \nu \lambda }^{*}W\right. 
\nonumber \\
&&\left. +3g\left( H_{\mu }^{*}H_{\nu }^{*}\frac{\delta ^{2}W}{\delta
\varphi ^{2}}A_{\lambda }+2H_{\mu }^{*}B_{\nu \lambda }^{*}\frac{\delta W}{%
\delta \varphi }\right) \right) C^{\mu \nu \lambda }  \nonumber \\
&&+\left( 4C_{\mu \nu \rho }^{*}\left( \partial _{\lambda }+g\frac{\delta W}{%
\delta \varphi }A_{\lambda }\right) +2g\eta _{\mu \nu \rho \lambda
}^{*}W-8g\eta _{\mu \nu \rho }^{*}H_{\lambda }^{*}\frac{\delta W}{\delta
\varphi }\right.  \nonumber \\
&&+4g\left( C_{\left[ \mu \nu \right. }^{*}H_{\left. \rho \right] }^{*}\frac{%
\delta ^{2}W}{\delta \varphi ^{2}}+H_{\mu }^{*}H_{\nu }^{*}H_{\rho }^{*}%
\frac{\delta ^{3}W}{\delta \varphi ^{3}}\right) A_{\lambda }  \nonumber \\
&&\left. +12g\left( C_{\mu \nu }^{*}\frac{\delta W}{\delta \varphi }+H_{\mu
}^{*}H_{\nu }^{*}\frac{\delta ^{2}W}{\delta \varphi ^{2}}\right) B_{\rho
\lambda }^{*}\right) C^{\mu \nu \rho \lambda }  \nonumber \\
&&+g\left( C_{\mu \nu }^{*}\frac{\delta W}{\delta \varphi }+H_{\mu
}^{*}H_{\nu }^{*}\frac{\delta ^{2}W}{\delta \varphi ^{2}}\right) \eta C^{\mu
\nu }  \nonumber \\
&&-g\left( C_{\mu \nu \rho }^{*}\frac{\delta W}{\delta \varphi }+C_{\left[
\mu \nu \right. }^{*}H_{\left. \rho \right] }^{*}\frac{\delta ^{2}W}{\delta
\varphi ^{2}}+H_{\mu }^{*}H_{\nu }^{*}H_{\rho }^{*}\frac{\delta ^{3}W}{%
\delta \varphi ^{3}}\right) \eta C^{\mu \nu \rho }  \nonumber \\
&&+g\left( C_{\mu \nu \rho \lambda }^{*}\frac{\delta W}{\delta \varphi }%
+\left( H_{\left[ \mu \right. }^{*}C_{\left. \nu \rho \lambda \right]
}^{*}+C_{\left[ \mu \nu \right. }^{*}C_{\left. \rho \lambda \right]
}^{*}\right) \frac{\delta ^{2}W}{\delta \varphi ^{2}}\right.  \nonumber \\
&&\left. \left. +H_{\left[ \mu \right. }^{*}H_{\nu }^{*}C_{\left. \rho
\lambda \right] }^{*}\frac{\delta ^{3}W}{\delta \varphi ^{3}}+H_{\mu
}^{*}H_{\nu }^{*}H_{\rho }^{*}H_{\lambda }^{*}\frac{\delta ^{4}W}{\delta
\varphi ^{4}}\right) \eta C^{\mu \nu \rho \lambda }\right) .  \label{bfa47}
\end{eqnarray}

By virtue of the discussion from the end of the subsection 3.1 on the
significance of terms with various antighost numbers in the solution to the
master equation, at this stage we can extract information on the gauge
structure of the coupled model. From the antifield-independent piece in (\ref
{bfa47}) we read that the overall Lagrangian action of the interacting gauge
theory has the expression 
\begin{eqnarray}
&&\tilde{S}[A^{\mu },H^{\mu },\varphi ,B^{\mu \nu },y^{i}]=\int d^{4}x\left(
H_{\mu }\left( \partial ^{\mu }\varphi -gW\left( \varphi \right) A^{\mu
}\right) \right.  \nonumber \\
&&\left. +\frac{1}{2}B^{\mu \nu }\partial _{[\mu }A_{\nu ]}+gU\left( \varphi
\right) j^{\mu }\left( \left[ y^{j}\right] \right) A_{\mu }+\mathcal{L}_{0}^{%
\mathrm{matt}}\left( y^{i},\partial _{\mu }y^{i},\cdots \right) \right) ,
\label{bfa48}
\end{eqnarray}
while from the components linear in the antighost number one antifields we
conclude that it is invariant under the gauge transformations 
\begin{equation}
\bar{\delta}_{\epsilon }A^{\mu }=\partial ^{\mu }\epsilon ,\;\bar{\delta}%
_{\epsilon }H^{\mu }=2D_{\nu }\epsilon ^{\mu \nu }+g\left( \frac{\delta U}{%
\delta \varphi }j^{\mu }\left( \left[ y^{j}\right] \right) -\frac{\delta W}{%
\delta \varphi }H^{\mu }\right) \epsilon ,  \label{bfa49}
\end{equation}
\begin{equation}
\bar{\delta}_{\epsilon }\varphi =gW\left( \varphi \right) \epsilon ,\;\bar{%
\delta}_{\epsilon }B^{\mu \nu }=-3\partial _{\rho }\epsilon ^{\mu \nu \rho
}+2gW\left( \varphi \right) \epsilon ^{\mu \nu },  \label{bfa49abc}
\end{equation}
\begin{equation}
\bar{\delta}_{\epsilon }y^{i}=gU\left( \varphi \right) T^{i}\left( \left[
y^{j}\right] \right) \epsilon ,  \label{bfa50}
\end{equation}
where we employed the notation 
\begin{equation}
D_{\nu }=\partial _{\nu }+g\frac{\delta W}{\delta \varphi }A_{\nu }.
\label{bfa49a}
\end{equation}
We remark on the one hand that the interaction term $U\left( \varphi \right)
j^{\mu }\left( \left[ y^{j}\right] \right) A_{\mu }$ expresses a generalized
minimal coupling and on the other hand that the gauge transformation of the
one-form $A_{\mu }$ from the BF-like theory remains $U\left( 1\right) $%
-abelian. From (\ref{bfa49}--\ref{bfa50}), we read the new non-vanishing
gauge generators 
\begin{equation}
(\tilde{Z}_{(A)}^{\mu })(x,x^{\prime })=(Z_{(A)}^{\mu })(x,x^{\prime
})=\partial _{x}^{\mu }\delta ^{4}(x-x^{\prime }),  \label{bfa50a}
\end{equation}
\begin{equation}
(\tilde{Z}_{(H)}^{\mu })_{\alpha \beta }(x,x^{\prime })=-D_{\left[ \alpha
\right. }^{x}\delta _{\left. \beta \right] }^{\mu }\delta ^{4}(x-x^{\prime
}),  \label{bfa50b}
\end{equation}
\begin{equation}
(\tilde{Z}_{(H)}^{\mu })(x,x^{\prime })=g\left( \frac{\delta U}{\delta
\varphi }\left( x\right) j^{\mu }\left( \left[ y^{j}\left( x\right) \right]
\right) -\frac{\delta W}{\delta \varphi }\left( x\right) H^{\mu }\left(
x\right) \right) \delta ^{4}(x-x^{\prime }),  \label{bfa50c}
\end{equation}
\begin{equation}
(\tilde{Z}_{(\varphi )})(x,x^{\prime })=gW\left( \varphi \left( x\right)
\right) \delta ^{4}(x-x^{\prime }),  \label{bfa50d}
\end{equation}
\begin{equation}
(\tilde{Z}_{(B)}^{\mu \nu })_{\alpha \beta \gamma }(x,x^{\prime
})=(Z_{(B)}^{\mu \nu })_{\alpha \beta \gamma }(x,x^{\prime })=-\frac{1}{2}%
\partial _{\left[ \alpha \right. }^{x}\delta _{\beta }^{\mu }\delta _{\left.
\gamma \right] }^{\nu }\delta ^{4}(x-x^{\prime }),  \label{bfa50g}
\end{equation}
\begin{equation}
(\tilde{Z}_{(B)}^{\mu \nu })_{\alpha \beta }(x,x^{\prime })=gW\left( \varphi
\left( x\right) \right) \delta _{\left[ \alpha \right. }^{\mu }\delta
_{\left. \beta \right] }^{\nu }\delta ^{4}(x-x^{\prime }),  \label{bfa50e}
\end{equation}
\begin{equation}
(\tilde{Z}_{(y)}^{i})(x,x^{\prime })=gU\left( \varphi \left( x\right)
\right) T^{i}\left( \left[ y^{j}\left( x\right) \right] \right) \delta
^{4}(x-x^{\prime }).  \label{bfa50f}
\end{equation}
The presence of the terms linear in the ghosts with pure ghost number two
and three in (\ref{bfa47}) shows that the gauge generators of the coupled
model are also second-order reducible, but some of the reducibility
functions are modified and, moreover, some of the reducibility relations
only hold on-shell, where on-shell means on the stationary surface of field
equations for the action (\ref{bfa48}). From the analysis of these terms we
infer the first-order reducibility functions 
\begin{equation}
(\tilde{Z}_{1}^{\alpha \beta })_{\mu ^{\prime }\nu ^{\prime }\rho ^{\prime
}}(x,x^{\prime })=-\frac{1}{2}D_{\left[ \mu ^{\prime }\right. }^{x}\delta
_{\nu ^{\prime }}^{\alpha }\delta _{\left. \rho ^{\prime }\right] }^{\beta
}\delta ^{4}(x-x^{\prime }),  \label{bfax1}
\end{equation}
\begin{equation}
(\tilde{Z}_{1}^{\alpha \beta \gamma })_{\mu ^{\prime }\nu ^{\prime }\rho
^{\prime }}(x,x^{\prime })=-\frac{1}{3}gW\left( \varphi \left( x\right)
\right) \delta _{\left[ \mu ^{\prime }\right. }^{\alpha }\delta _{\nu
^{\prime }}^{\beta }\delta _{\left. \rho ^{\prime }\right] }^{\gamma }\delta
^{4}(x-x^{\prime }),  \label{bfax2}
\end{equation}
\begin{eqnarray}
&&(\tilde{Z}_{1}^{\alpha \beta \gamma })_{\mu ^{\prime }\nu ^{\prime }\rho
^{\prime }\lambda ^{\prime }}(x,x^{\prime })=(Z_{1}^{\alpha \beta \gamma
})_{\mu ^{\prime }\nu ^{\prime }\rho ^{\prime }\lambda ^{\prime
}}(x,x^{\prime })=  \nonumber \\
&&-\frac{1}{6}\partial _{\left[ \mu ^{\prime }\right. }^{x}\delta _{\nu
^{\prime }}^{\alpha }\delta _{\rho ^{\prime }}^{\beta }\delta _{\left.
\lambda ^{\prime }\right] }^{\gamma }\delta ^{4}(x-x^{\prime }),
\label{bfax3}
\end{eqnarray}
and also the second-order ones 
\begin{equation}
\left( \tilde{Z}_{2}^{\mu ^{\prime }\nu ^{\prime }\rho ^{\prime }}\right)
_{\alpha ^{\prime }\beta ^{\prime }\gamma ^{\prime }\delta ^{\prime
}}(x,x^{\prime })=-\frac{1}{6}D_{\left[ \alpha ^{\prime }\right. }^{x}\delta
_{\beta ^{\prime }}^{\mu ^{\prime }}\delta _{\gamma ^{\prime }}^{\nu
^{\prime }}\delta _{\left. \delta ^{\prime }\right] }^{\rho ^{\prime
}}\delta ^{4}(x-x^{\prime }),  \label{bfax4}
\end{equation}
\begin{equation}
(\tilde{Z}_{2}^{\mu ^{\prime }\nu ^{\prime }\rho ^{\prime }\lambda ^{\prime
}})_{\alpha ^{\prime }\beta ^{\prime }\gamma ^{\prime }\delta ^{\prime
}}(x,x^{\prime })=\frac{1}{12}gW\left( \varphi \left( x\right) \right)
\delta _{\left[ \alpha ^{\prime }\right. }^{\mu ^{\prime }}\delta _{\beta
^{\prime }}^{\nu ^{\prime }}\delta _{\gamma ^{\prime }}^{\rho ^{\prime
}}\delta _{\left. \delta ^{\prime }\right] }^{\lambda ^{\prime }}\delta
^{4}(x-x^{\prime }),  \label{bfax5}
\end{equation}
as well as the first- and second-order reducibility relations (written in De
Witt condensed notations) 
\begin{eqnarray}
&&(\tilde{Z}_{(H)}^{\mu })_{\alpha \beta }(\tilde{Z}_{1}^{\alpha \beta
})_{\mu ^{\prime }\nu ^{\prime }\rho ^{\prime }}=-g\frac{\delta ^{2}W}{%
\delta \varphi ^{2}}A_{\left[ \mu ^{\prime }\right. }\delta _{\nu ^{\prime
}}^{\mu }\frac{\delta \tilde{S}}{\delta H^{\left. \rho ^{\prime }\right] }} 
\nonumber \\
&&-2g\frac{\delta W}{\delta \varphi }\delta _{\left[ \mu ^{\prime }\right.
}^{\mu }\frac{\delta \tilde{S}}{\delta B^{\left. \nu ^{\prime }\rho ^{\prime
}\right] }},  \label{bfax6}
\end{eqnarray}
\begin{eqnarray}
&&(\tilde{Z}_{(B)}^{\mu \nu })_{\alpha \beta }(\tilde{Z}_{1}^{\alpha \beta
})_{\mu ^{\prime }\nu ^{\prime }\rho ^{\prime }}+(\tilde{Z}_{(B)}^{\mu \nu
})_{\alpha \beta \gamma }(\tilde{Z}_{1}^{\alpha \beta \gamma })_{\mu
^{\prime }\nu ^{\prime }\rho ^{\prime }}=  \nonumber \\
&&g\frac{\delta W}{\delta \varphi }\delta _{\left[ \mu ^{\prime }\right.
}^{\mu }\delta _{\nu ^{\prime }}^{\nu }\frac{\delta \tilde{S}}{\delta
H^{\left. \rho ^{\prime }\right] }},  \label{bfax7}
\end{eqnarray}
\begin{equation}
(\tilde{Z}_{(B)}^{\mu \nu })_{\alpha \beta \gamma }(\tilde{Z}^{\alpha \beta
\gamma })_{\mu ^{\prime }\nu ^{\prime }\rho ^{\prime }\lambda ^{\prime }}=0,
\label{bfax8}
\end{equation}
\begin{eqnarray}
&&(\tilde{Z}_{1}^{\alpha \beta })_{\mu ^{\prime }\nu ^{\prime }\rho ^{\prime
}}\left( \tilde{Z}_{2}^{\mu ^{\prime }\nu ^{\prime }\rho ^{\prime }}\right)
_{\alpha ^{\prime }\beta ^{\prime }\gamma ^{\prime }\delta ^{\prime }}=\frac{%
g}{2}\frac{\delta ^{2}W}{\delta \varphi ^{2}}A_{\left[ \alpha ^{\prime
}\right. }\delta _{\beta ^{\prime }}^{\alpha }\delta _{\gamma ^{\prime
}}^{\beta }\frac{\delta \tilde{S}}{\delta H^{\left. \delta ^{\prime }\right]
}}  \nonumber \\
&&-g\frac{\delta W}{\delta \varphi }\delta _{\left[ \alpha ^{\prime }\right.
}^{\alpha }\delta _{\beta ^{\prime }}^{\beta }\frac{\delta \tilde{S}}{\delta
B^{\left. \gamma ^{\prime }\delta ^{\prime }\right] }},  \label{bfax9}
\end{eqnarray}
\begin{eqnarray}
&&(\tilde{Z}_{1}^{\alpha \beta \gamma })_{\mu ^{\prime }\nu ^{\prime }\rho
^{\prime }}\left( \tilde{Z}_{2}^{\mu ^{\prime }\nu ^{\prime }\rho ^{\prime
}}\right) _{\alpha ^{\prime }\beta ^{\prime }\gamma ^{\prime }\delta
^{\prime }}+(\tilde{Z}_{1}^{\alpha \beta \gamma })_{\mu ^{\prime }\nu
^{\prime }\rho ^{\prime }\lambda ^{\prime }}(\tilde{Z}_{2}^{\mu ^{\prime
}\nu ^{\prime }\rho ^{\prime }\lambda ^{\prime }})_{\alpha ^{\prime }\beta
^{\prime }\gamma ^{\prime }\delta ^{\prime }}=  \nonumber \\
&&\frac{g}{3}\frac{\delta W}{\delta \varphi }\delta _{\left[ \alpha ^{\prime
}\right. }^{\alpha }\delta _{\beta ^{\prime }}^{\beta }\delta _{\gamma
^{\prime }}^{\gamma }\frac{\delta \tilde{S}}{\delta H^{\left. \delta
^{\prime }\right] }}.  \label{bfax10}
\end{eqnarray}
The pieces from (\ref{bfa47}) that are quadratic in the ghosts of pure ghost
number one are of two kinds: ones are linear in their antifields, and the
others are quadratic in the antifields of the original fields, such that the
deformed gauge algebra is open. The non-vanishing commutators among the
gauge transformations of the coupled model read as (again in De Witt
condensed notations) 
\begin{eqnarray}
&&(\tilde{Z}_{(\varphi )})\frac{\delta (\tilde{Z}_{(H)}^{\mu })_{\alpha
\beta }}{\delta \varphi }+(\tilde{Z}_{(A)}^{\rho })\frac{\delta (\tilde{Z}%
_{(H)}^{\mu })_{\alpha \beta }}{\delta A^{\rho }}-(\tilde{Z}_{(H)}^{\rho
})_{\alpha \beta }\frac{\delta (\tilde{Z}_{(H)}^{\mu })}{\delta H^{\rho }}= 
\nonumber \\
&&-g\frac{\delta W}{\delta \varphi }(\tilde{Z}_{(H)}^{\mu })_{\alpha \beta
}+g\frac{\delta ^{2}W}{\delta \varphi ^{2}}\delta _{\left[ \alpha \right.
}^{\mu }\frac{\delta \tilde{S}}{\delta H^{\left. \beta \right] }},
\label{bfax11}
\end{eqnarray}
\begin{equation}
(\tilde{Z}_{(\varphi )})\frac{\delta (\tilde{Z}_{(B)}^{\mu \nu })_{\alpha
\beta }}{\delta \varphi }=g\frac{\delta W}{\delta \varphi }(\tilde{Z}%
_{(B)}^{\mu \nu })_{\alpha \beta }.  \label{bfax12}
\end{equation}
The remaining elements in (\ref{bfa47}) give us information on the
higher-order gauge structure of the interacting model.

In conclusion, if the conserved current present in the purely matter theory
is still invariant under the gauge version of the initial global
one-parameter symmetry, then the deformation procedure stops at order one in
the coupling constant and induces the gauging of the generalized rigid
symmetry (see the expression of the functions $\bar{T}^{i}$ from (\ref
{bfa39a})) at the level of the matter fields as in formula (\ref{bfa50}).
Moreover, the one-form $H^{\mu }$ from the BF-sector gains one-parameter
gauge transformations whose generators involve the conserved currents of the
matter theory and the first-order derivative of the `background' potential.
In addition, the interaction couples the matter fields to the one-form $%
A^{\mu }$ from the BF-like theory only at the first order in the deformation
parameter through the gauge invariant matter current and the `background'
potential by means of a generalized minimal coupling.

(I.2) In the opposite situation, where (\ref{ide13}) is satisfied, but the
matter current is not invariant under the gauge version of the initial
global one-parameter symmetry (\ref{bfa40}) 
\begin{equation}
\frac{\delta ^{R}j^{\mu }\left( \left[ y^{j}\right] \right) }{\delta y^{i}}%
T^{i}\left( \left[ y^{k}\right] \right) \neq 0,  \label{bfa51}
\end{equation}
it follows that $\left( S_{1},S_{1}\right) $ is non-vanishing (see the right
hand-side of the equation (\ref{ide14})), hence the second-order deformation 
$S_{2}$ involved with the equation (\ref{bfa2.6}) will also be so. Moreover,
it is possible to obtain other non-trivial higher-order deformations when
solving the remaining equations ((\ref{bfa2.7}), etc.). Nevertheless, the
expressions of these deformations strongly depend on the structure of the
matter theory and cannot be output in the general setting considered here.
However, we expect that the consistency equations (\ref{bfa2.6}--\ref{bfa2.7}%
), etc. impose further restrictions on the function $U\left( \varphi \right) 
$. What is always valid is that the complete deformed solution to the master
equation starts like 
\begin{equation}
S^{^{\prime }}=S+g^{2}S_{2}+\cdots ,  \label{bfa52}
\end{equation}
such that the Lagrangian action of the interacting theory is of the type 
\begin{equation}
\tilde{S}^{^{\prime }}=\tilde{S}+\mathcal{O}\left( g^{2}\right) ,
\label{bfa53}
\end{equation}
where $S$ and $\tilde{S}$ are expressed by (\ref{bfa47}) and (\ref{bfa48})
respectively.

(II) In the complementary situation, where 
\begin{equation}
W\left( \varphi \right) =0,  \label{ide15}
\end{equation}
and $M\left( \varphi \right) $ is an arbitrary function of the
undifferentiated scalar field, from (\ref{bfa45}) it follows that 
\begin{eqnarray}
&&\frac{1}{2}\left( S_{1},S_{1}\right) =-2\varepsilon _{\mu \nu \rho \lambda
}\int d^{4}x\left( \left( \frac{\delta \left( MU\right) }{\delta \varphi }%
j^{\alpha }H_{\alpha }^{*}+MUy_{i}^{*}T^{i}\right) \eta ^{\mu \nu \rho
\lambda }\right.  \nonumber \\
&&\left. +MUj^{\mu }\eta ^{\nu \rho \lambda }\right) +\int d^{4}xU^{2}\frac{%
\delta ^{R}j^{\mu }}{\delta y^{i}}T^{i}A_{\mu }\eta .  \label{bfa145}
\end{eqnarray}
Nevertheless, according to the local version of the equation (\ref{bfa2.6}),
the non-integrated density from the right hand-side of (\ref{bfa145}) must
again be written like the $s$-variation modulo $d$ of a local function.
Taking into account the actions (\ref{bfa15}--\ref{bfa23}) of $\gamma $ and $%
\delta $ on the BRST generators, it results that the piece in (\ref{bfa145})
proportional with $\varepsilon _{\mu \nu \rho \lambda }$ is $s$-exact if and
only if the matter current $j^{\mu }$ is trivial (see the relations (\ref
{rigidm}) and (\ref{trivy})). Indeed, in this situation direct computation
leads to 
\begin{eqnarray}
&&-2\varepsilon _{\mu \nu \rho \lambda }\left( \left( \frac{\delta \left(
MU\right) }{\delta \varphi }j^{\alpha }H_{\alpha
}^{*}+MUy_{i}^{*}T^{i}\right) \eta ^{\mu \nu \rho \lambda }+MUj^{\mu }\eta
^{\nu \rho \lambda }\right) \rightarrow  \nonumber \\
&&-2\varepsilon _{\mu \nu \rho \lambda }\left( \frac{\delta \left( MU\right) 
}{\delta \varphi }H_{\alpha }^{*}\partial _{\beta }t^{\alpha \beta
}+MUy_{i}^{*}M^{ij}\frac{\delta ^{L}\mathcal{L}_{0}^{\mathrm{matt}}}{\delta
y^{j}}\right) \eta ^{\mu \nu \rho \lambda }  \nonumber \\
&&-2\varepsilon _{\mu \nu \rho \lambda }\left( MU\right) \eta ^{\nu \rho
\lambda }\partial _{\beta }t^{\mu \beta }=-\varepsilon _{\mu \nu \rho 
\lambda }s\left( \left( \left( \frac{\delta \left( MU\right) }{\delta \varphi }%
C_{\alpha \beta }^{*}\right. \right. \right.  \nonumber \\
&&\left. \left. +\frac{\delta ^{2}\left( MU\right) }{\delta \varphi ^{2}}%
H_{\alpha }^{*}H_{\beta }^{*}\right) t^{\alpha \beta
}-MUy_{i}^{*}M^{ij}y_{j}^{*}\right) \eta ^{\mu \nu \rho \lambda }  \nonumber
\\
&&\left. +2\frac{\delta \left( MU\right) }{\delta \varphi }H_{\alpha
}^{*}t^{\alpha \mu }\eta ^{\nu \rho \lambda }+\left( MU\right) t^{\mu \nu
}B^{\rho \lambda }\right)  \nonumber \\
&&+\partial _{\beta }\left( 2\varepsilon _{\mu \nu \rho \lambda }\left( 
\frac{\delta \left( MU\right) }{\delta \varphi }t^{\beta \alpha }H_{\alpha
}^{*}\eta ^{\mu \nu \rho \lambda }+MUt^{\beta \mu }\eta ^{\nu \rho \lambda
}\right) \right) .  \label{bfa145abc}
\end{eqnarray}
As discussed in the above (see the formula (\ref{rigidkn})), in this case
the terms $a_{1}^{\mathrm{(BF-m)}}+a_{0}^{\mathrm{(BF-m)}}$ can be
completely removed from the first-order deformation via the transformation 
\begin{eqnarray}
S_{1} &\rightarrow &S_{1}^{\prime }=S_{1}+s\int d^{4}x\left( t^{\mu \nu
}\left( \left[ y^{j}\right] \right) \left( \frac{1}{2}\frac{\delta
^{2}U\left( \varphi \right) }{\delta \varphi ^{2}}H_{\mu }^{*}H_{\nu
}^{*}\eta \right. \right.  \nonumber \\
&&\left. +\frac{\delta U\left( \varphi \right) }{\delta \varphi }\left( 
\frac{1}{2}C_{\mu \nu }^{*}\eta +H_{\mu }^{*}A_{\nu }\right) +B_{\mu \nu
}^{*}U\left( \varphi \right) \right)  \nonumber \\
&&\left. -\frac{1}{2}U\left( \varphi \right) y_{i}^{*}M^{ij}\left( \left[
y^{k}\right] \right) y_{j}^{*}\eta \right) ,  \label{nous1}
\end{eqnarray}
and thus we find that 
\begin{equation}
\frac{1}{2}\left( S_{1}^{\prime },S_{1}^{\prime }\right) =0,  \label{higher}
\end{equation}
such that we can safely take the second-order deformation equal to zero, $%
S_{2}=0$. The higher-order consistency equations are then fulfilled with the
choice $S_{3}=S_{4}=\cdots =0$. Thus, if (\ref{ide15}) is satisfied, then
the complete deformed solution to the master equation, which is consistent
at all orders in the coupling constant, is given by 
\begin{eqnarray}
&&S=\int d^{4}x\left( H_{\mu }\partial ^{\mu }\varphi +\frac{1}{2}B^{\mu \nu
}\partial _{[\mu }A_{\nu ]}+\mathcal{L}_{0}^{\mathrm{matt}}\left(
y^{i},\partial _{\mu }y^{i},\cdots \right) \right.  \nonumber \\
&&+\frac{g}{2}M\varepsilon _{\alpha \beta \gamma \delta }B^{\alpha \beta
}B^{\gamma \delta }+A^{*\mu }\left( \partial _{\mu }\eta -2gM\varepsilon
_{\mu \alpha \beta \gamma }\eta ^{\alpha \beta \gamma }\right)  \nonumber \\
&&+2H_{\mu }^{*}\left( \partial _{\nu }C^{\mu \nu }+g\frac{\delta M}{\delta
\varphi }B^{\mu \alpha }\varepsilon _{\alpha \beta \gamma \delta }\eta
^{\beta \gamma \delta }\right) -3B_{\mu \nu }^{*}\partial _{\rho }\eta ^{\mu
\nu \rho }  \nonumber \\
&&-3C_{\mu \nu }^{*}\partial _{\rho }C^{\mu \nu \rho }+4\eta _{\mu \nu \rho
}^{*}\partial _{\lambda }\eta ^{\mu \nu \rho \lambda }  \nonumber \\
&&+g\left( \left( \frac{\delta M}{\delta \varphi }C_{\rho \lambda }^{*}+%
\frac{\delta ^{2}M}{\delta \varphi ^{2}}H_{\rho }^{*}H_{\lambda }^{*}\right)
B^{\rho \lambda }+2\left( \frac{\delta M}{\delta \varphi }H_{\mu
}^{*}A^{*\mu }-M\eta ^{*}\right) \right) \varepsilon _{\alpha \beta \gamma
\delta }\eta ^{\alpha \beta \gamma \delta }  \nonumber \\
&&-\frac{9}{4}g\left( \frac{\delta M}{\delta \varphi }C_{\rho \lambda }^{*}+%
\frac{\delta ^{2}M}{\delta \varphi ^{2}}H_{\rho }^{*}H_{\lambda }^{*}\right)
\varepsilon _{\alpha \beta \gamma \delta }\eta ^{\rho \alpha \beta }\eta
^{\lambda \gamma \delta }+4C_{\mu \nu \rho }^{*}\partial _{\lambda }C^{\mu
\nu \rho \lambda }  \nonumber \\
&&-g\left( \frac{\delta M}{\delta \varphi }C_{\nu \rho \lambda }^{*}+\frac{%
\delta ^{2}M}{\delta \varphi ^{2}}H_{\left[ \nu \right. }^{*}C_{\left. \rho
\lambda \right] }^{*}+\frac{\delta ^{3}M}{\delta \varphi ^{3}}H_{\nu
}^{*}H_{\rho }^{*}H_{\lambda }^{*}\right) \eta ^{\nu \rho \lambda
}\varepsilon _{\alpha \beta \gamma \delta }\eta ^{\alpha \beta \gamma \delta
}  \nonumber \\
&&+\frac{g}{2}\left( \frac{\delta M}{\delta \varphi }C_{\mu \nu \rho \lambda
}^{*}+\frac{\delta ^{2}M}{\delta \varphi ^{2}}H_{\left[ \mu \right.
}^{*}C_{\left. \nu \rho \lambda \right] }^{*}+\frac{\delta ^{2}M}{\delta
\varphi ^{2}}C_{\left[ \mu \nu \right. }^{*}C_{\left. \rho \lambda \right]
}^{*}\right.  \nonumber \\
&&\left. \left. +\frac{\delta ^{3}M}{\delta \varphi ^{3}}H_{\left[ \mu
\right. }^{*}H_{\nu }^{*}C_{\left. \rho \lambda \right] }^{*}+\frac{\delta
^{4}M}{\delta \varphi ^{4}}H_{\mu }^{*}H_{\nu }^{*}H_{\rho }^{*}H_{\lambda
}^{*}\right) \eta ^{\mu \nu \rho \lambda }\varepsilon _{\alpha \beta \gamma
\delta }\eta ^{\alpha \beta \gamma \delta }\right) .  \label{defcaseii}
\end{eqnarray}

Its antighost number zero part emphasizes the Lagrangian action of the
deformed theory 
\begin{eqnarray}
&&\tilde{S}[A^{\mu },H^{\mu },\varphi ,B^{\mu \nu },y^{i}]=\int d^{4}x\left( 
\mathcal{L}_{0}^{\mathrm{matt}}\left( y^{i},\partial _{\mu }y^{i},\cdots
\right) \right.  \nonumber \\
&&\left. +H_{\mu }\partial ^{\mu }\varphi +\frac{1}{2}B^{\mu \nu }\partial
_{[\mu }A_{\nu ]}+\frac{g}{2}M\left( \varphi \right) \varepsilon _{\alpha
\beta \gamma \delta }B^{\alpha \beta }B^{\gamma \delta }\right) ,
\label{sdefcaseii}
\end{eqnarray}
while the antighost number one components provide the gauge transformations
of the action (\ref{sdefcaseii}) 
\begin{equation}
\bar{\delta}_{\epsilon }A_{\mu }=\partial _{\mu }\epsilon -2gM\left( \varphi
\right) \varepsilon _{\mu \alpha \beta \gamma }\epsilon ^{\alpha \beta
\gamma }\equiv (\tilde{Z}_{(A)\mu })\epsilon +(\tilde{Z}_{(A)\mu })_{\alpha
\beta \gamma }\epsilon ^{\alpha \beta \gamma },  \label{defg2}
\end{equation}
\begin{equation}
\bar{\delta}_{\epsilon }H^{\mu }=2\left( \partial _{\nu }\epsilon ^{\mu \nu
}-g\frac{\delta M}{\delta \varphi }B^{\mu \alpha }\varepsilon _{\alpha \beta
\gamma \delta }\epsilon ^{\beta \gamma \delta }\right) \equiv (\tilde{Z}%
_{(H)}^{\mu })_{\alpha \beta }\epsilon ^{\alpha \beta }+(\tilde{Z}%
_{(H)}^{\mu })_{\alpha \beta \gamma }\epsilon ^{\alpha \beta \gamma },
\label{defgg2}
\end{equation}
\begin{equation}
\bar{\delta}_{\epsilon }\varphi =0,\;\bar{\delta}_{\epsilon }B^{\mu \nu
}=-3\partial _{\rho }\epsilon ^{\mu \nu \rho }\equiv (\tilde{Z}_{(B)}^{\mu
\nu })_{\alpha \beta \gamma }\epsilon ^{\alpha \beta \gamma },\;\bar{\delta}%
_{\epsilon }y^{i}=0.  \label{deffg2}
\end{equation}
We observe that in this case the matter fields remain uncoupled to the BF
field sector. From (\ref{defg2}--\ref{deffg2}), we notice that the
non-vanishing gauge generators are 
\begin{equation}
(\tilde{Z}_{(A)\mu })(x,x^{\prime })=(Z_{(A)\mu })(x,x^{\prime })=\partial
_{\mu }^{x}\delta ^{4}(x-x^{\prime }),  \label{vechiA}
\end{equation}
\begin{equation}
(\tilde{Z}_{(A)\mu })_{\alpha \beta \gamma }(x,x^{\prime })=-2gM\left(
\varphi \left( x\right) \right) \varepsilon _{\mu \alpha \beta \gamma
}\delta ^{4}(x-x^{\prime }),  \label{noiA}
\end{equation}
\begin{equation}
(\tilde{Z}_{(H)}^{\mu })_{\alpha \beta }(x,x^{\prime })=(Z_{(H)}^{\mu
})_{\alpha \beta }(x,x^{\prime })=-\partial _{\left[ \alpha \right.
}^{x}\delta _{\left. \beta \right] }^{\mu }\delta ^{4}(x-x^{\prime }),
\label{vechiH}
\end{equation}
\begin{equation}
(\tilde{Z}_{(H)}^{\mu })_{\alpha \beta \gamma }(x,x^{\prime })=-2g\frac{%
\delta M}{\delta \varphi }\left( x\right) B^{\mu \nu }\left( x\right)
\varepsilon _{\nu \alpha \beta \gamma }\delta ^{4}(x-x^{\prime }),
\label{noiH}
\end{equation}
\begin{equation}
(\tilde{Z}_{(B)}^{\mu \nu })_{\alpha \beta \gamma }(x,x^{\prime
})=(Z_{(B)}^{\mu \nu })_{\alpha \beta \gamma }(x,x^{\prime })=-\frac{1}{2}%
\partial _{\left[ \alpha \right. }^{x}\delta _{\beta }^{\mu }\delta _{\left.
\gamma \right] }^{\nu }\delta ^{4}(x-x^{\prime }).  \label{vechiB}
\end{equation}
Thus, the scalar field and the matter fields are still not endowed with
gauge transformations, while the gauge transformations of the two-form are
also not changed with respect to the ``free'' model. Actually, the gauge
transformations become richer than in the ``free'' case only for the
one-forms $A^{\mu }$ and $H^{\mu }$ in the sector associated with the gauge
parameters $\epsilon ^{\alpha \beta \gamma }$. The deformed gauge algebra
(corresponding to the generating set (\ref{defg2}--\ref{deffg2})) is open,
as can be seen from the elements of antighost number two in (\ref{defcaseii}%
) that are quadratic in the ghosts of pure ghost number one. The only
non-abelian commutators among the new gauge transformations are expressed by 
\begin{eqnarray}
&&\left( \tilde{Z}_{(B)}^{\rho \lambda }\right) _{\alpha \beta \gamma }\frac{%
\delta (\tilde{Z}_{(H)}^{\mu })_{\alpha ^{\prime }\beta ^{\prime }\gamma
^{\prime }}}{\delta B^{\rho \lambda }}-\left( \tilde{Z}_{(B)}^{\rho \lambda
}\right) _{\alpha ^{\prime }\beta ^{\prime }\gamma ^{\prime }}\frac{\delta (%
\tilde{Z}_{(H)}^{\mu })_{\alpha \beta \gamma }}{\delta B^{\rho \lambda }}= 
\nonumber \\
&&-\frac{g}{4}\frac{\delta M}{\delta \varphi }(\tilde{Z}_{(H)}^{\mu })_{\rho
\lambda }\delta _{\left[ \alpha \right. }^{\left[ \rho \right. }\left(
\varepsilon _{\left. \beta \gamma \right] \left[ \alpha ^{\prime }\beta
^{\prime }\right. }\right) \delta _{\left. \gamma ^{\prime }\right]
}^{\left. \lambda \right] }  \nonumber \\
&&+g\frac{\delta ^{2}M}{\delta \varphi ^{2}}\delta _{\left[ \alpha \right.
}^{\left[ \mu \right. }\left( \varepsilon _{\left. \beta \gamma \right]
\left[ \alpha ^{\prime }\beta ^{\prime }\right. }\right) \delta _{\left.
\gamma ^{\prime }\right] }^{\left. \nu \right] }\frac{\delta \tilde{S}}{%
\delta H^{\nu }}.  \label{algnoua}
\end{eqnarray}
Looking at the rest of the terms with antighost number two in (\ref
{defcaseii}), we can state that, besides the original first-order
reducibility relation (\ref{bfa5a}), there appear some new ones 
\begin{equation}
(Z_{(H)}^{\mu })_{\rho \lambda }\left( \tilde{Z}_{1}^{\rho \lambda }\right)
_{\alpha \beta \gamma \delta }=2g\varepsilon _{\alpha \beta \gamma \delta
}\left( \frac{\delta ^{2}M}{\delta \varphi ^{2}}B^{\mu \nu }\frac{\delta 
\tilde{S}}{\delta H^{\nu }}+\frac{\delta M}{\delta \varphi }g^{\mu \nu }%
\frac{\delta \tilde{S}}{\delta A^{\nu }}\right) ,  \label{red1nou}
\end{equation}
\begin{equation}
(Z_{(A)}^{\mu })\left( \tilde{Z}_{1}\right) _{\alpha \beta \gamma \delta
}=-2g\varepsilon _{\alpha \beta \gamma \delta }\frac{\delta M}{\delta
\varphi }g^{\mu \nu }\frac{\delta \tilde{S}}{\delta H^{\nu }},
\label{red1noua}
\end{equation}
that only close on-shell (i.e., on the stationary surface of field equations
resulting from the action (\ref{sdefcaseii})), where the accompanying
first-order reducibility functions are of the form 
\begin{equation}
\left( \tilde{Z}_{1}^{\rho \lambda }\right) _{\alpha \beta \gamma \delta
}(x,x^{\prime })=g\frac{\delta M}{\delta \varphi }\left( x\right) B^{\rho
\lambda }\left( x\right) \varepsilon _{\alpha \beta \gamma \delta }\delta
^{4}(x-x^{\prime }),  \label{red1noi}
\end{equation}
\begin{equation}
\left( \tilde{Z}_{1}\right) _{\alpha \beta \gamma \delta }(x,x^{\prime
})=-2gM\left( \varphi \left( x\right) \right) \varepsilon _{\alpha \beta
\gamma \delta }\delta ^{4}(x-x^{\prime }).  \label{red1noii}
\end{equation}
The second-order reducibility is not modified (it continues to be expressed
by (\ref{bfa5b})). The presence in (\ref{defcaseii}) of elements with
antighost number strictly greater than two that are proportional with the
coupling constant $g$ signifies a higher-order gauge tensor structure of the
deformed model, due to the open character of the gauge algebra, as well as
to the field dependence of the deformed reducibility functions. The case
(II) appears thus to be less important from the perspective of constructing
effective couplings among the BF fields and the matter field sector, since
no non-trivial interactions among them are allowed.

This completes our general procedure.

\section{Applications}

Next, we consider two examples of matter theories, the massive complex
scalar field and the massive spin 3/2 field, and compute their consistent
interactions with the four-dimensional BF-like theory. Since we are merely
interested in the possibility of non-trivial couplings with the BF fields,
we restrict ourselves to the case (I) studied in the above, for which the
relation (\ref{ide13}) is assumed to hold.

\subsection{Complex scalar field}

In the sequel we apply the theoretical part of the paper to the case where
the matter theory describes a massive complex scalar field. In this
situation we have that $y^{i}=\left( \Phi ,\bar{\Phi}\right) $, where the
Lagrangian of the matter sector in (\ref{bfa1}) is expressed by 
\begin{equation}
\mathcal{L}_{0}^{\mathrm{matt}}=\left( \partial _{\mu }\Phi \right) \left(
\partial ^{\mu }\bar{\Phi}\right) -\mu ^{2}\Phi \bar{\Phi}-V\left( \Phi \bar{%
\Phi}\right) .  \label{bfa54}
\end{equation}
Here, the bar operation signifies complex conjugation. The Koszul-Tate
differential and the exterior longitudinal derivative composing the ``free''
BRST symmetry act on the generators from the matter sector $y^{i}=\left(
\Phi ,\bar{\Phi}\right) $ and on their antifields $y_{i}^{*}=\left( \Phi
^{*},\bar{\Phi}^{*}\right) $ like 
\begin{equation}
\delta \Phi =\delta \bar{\Phi}=0,\;\gamma \Phi =\gamma \bar{\Phi}=0,
\label{bfa55}
\end{equation}
\begin{equation}
\delta \Phi ^{*}\equiv -\frac{\delta \mathcal{L}_{0}^{\mathrm{matt}}}{\delta
\Phi }=\left( \partial _{\mu }\partial ^{\mu }+\mu ^{2}\right) \bar{\Phi}+%
\frac{\partial V}{\partial \left( \Phi \bar{\Phi}\right) }\bar{\Phi},
\label{bfa56}
\end{equation}
\begin{equation}
\delta \bar{\Phi}^{*}\equiv -\frac{\delta \mathcal{L}_{0}^{\mathrm{matt}}}{%
\delta \bar{\Phi}}=\left( \partial _{\mu }\partial ^{\mu }+\mu ^{2}\right)
\Phi +\frac{\partial V}{\partial \left( \Phi \bar{\Phi}\right) }\Phi ,
\label{bfa57}
\end{equation}
\begin{equation}
\gamma \Phi ^{*}=\gamma \bar{\Phi}^{*}=0.  \label{bfa58}
\end{equation}
We notice that the functions $\frac{\delta \mathcal{L}_{0}^{\mathrm{matt}}}{%
\delta \Phi }$ and $\frac{\delta \mathcal{L}_{0}^{\mathrm{matt}}}{\delta 
\bar{\Phi}}$ that define the field equations are independent and split into
two components: one is linear in the fields and second-order in their
spacetime derivatives, while the other, even if allowed to be non-linear,
does not involve the spacetime derivatives of the fields. Moreover, if we
denote by $I_{0}$ the set of independent field variables, then it is stable
under spatial differentiation ($\partial _{k}I_{0}\subset I_{0}$). Thus, the
complex scalar field theory is a normal theory without gauge invariance, of
Cauchy order one \cite{gen1}, in agreement with the hypothesis made in the
theoretical part of the paper. The actions of $\delta $ and $\gamma $ on the
fields/antifields pertaining to the BF sector can be found among the
definitions (\ref{bfa15}--\ref{bfa23}). Multiplying (\ref{bfa56}) by $\Phi $
and (\ref{bfa57}) by $\bar{\Phi}$, we arrive at the relation 
\begin{equation}
\delta \left[ i\left( \bar{\Phi}^{*}\bar{\Phi}-\Phi ^{*}\Phi \right) \right]
=\partial _{\mu }\left[ i\left( \bar{\Phi}\partial ^{\mu }\Phi -\Phi
\partial ^{\mu }\bar{\Phi}\right) \right] ,  \label{bfa59}
\end{equation}
which expresses the conservation of the non-trivial current 
\begin{equation}
j^{\mu }=i\left( \bar{\Phi}\partial ^{\mu }\Phi -\Phi \partial ^{\mu }\bar{%
\Phi}\right) ,  \label{bfa60}
\end{equation}
corresponding to the bosonic global one-parameter invariance 
\begin{equation}
\Delta \Phi =i\Phi \xi ,\Delta \bar{\Phi}=-i\bar{\Phi}\xi ,  \label{bfa61}
\end{equation}
of the complex scalar field action. By comparing (\ref{bfa61}) with (\ref
{bfa40}), it follows that 
\begin{equation}
T^{i}=\left( i\Phi ,-i\bar{\Phi}\right) .  \label{bfa62}
\end{equation}
Inserting (\ref{bfa60}) and (\ref{bfa62}) in (\ref{bfa44}) with $M\left(
\varphi \right) =0$, we then determine the first-order deformation of the
solution to the master equation, where 
\begin{eqnarray}
&&A_{\mu }U\left( \varphi \right) j^{\mu }\left( \left[ y^{j}\right] \right)
+\left( H_{\mu }^{*}\frac{\delta U}{\delta \varphi }j^{\mu }\left( \left[
y^{j}\right] \right) +y_{i}^{*}U\left( \varphi \right) T^{i}\left( \left[
y^{j}\right] \right) \right) \eta \rightarrow  \nonumber \\
&&i\left( A_{\mu }U\left( \varphi \right) \left( \bar{\Phi}\partial ^{\mu
}\Phi -\Phi \partial ^{\mu }\bar{\Phi}\right) \right.  \nonumber \\
&&\left. +\left( H_{\mu }^{*}\frac{\delta U}{\delta \varphi }\left( \bar{\Phi%
}\partial ^{\mu }\Phi -\Phi \partial ^{\mu }\bar{\Phi}\right) +U\left(
\varphi \right) \left( \Phi ^{*}\Phi -\bar{\Phi}^{*}\bar{\Phi}\right)
\right) \eta \right) .  \label{s1}
\end{eqnarray}

After some computation, we find that $\left( S_{1},S_{1}\right) $ is not
vanishing 
\begin{equation}
\left( S_{1},S_{1}\right) =4\int d^{4}xU^{2}\left( \varphi \right) \left(
\partial _{\mu }\left( \Phi \bar{\Phi}A^{\mu }\right) \right) \eta =\int
d^{4}x\Delta ,  \label{bfa63}
\end{equation}
due to the non-invariance of the current (\ref{bfa60}) under the gauge
version of the global invariance associated with the complex scalar theory.
We are precisely in the situation (I.2) analyzed in the previous section (it
is the equation (\ref{bfa51}) that applies now). Thus, we have to solve the
second-order deformation equation, expressed by 
\begin{equation}
\frac{1}{2}\Delta +sb=\partial _{\mu }\theta ^{\mu },  \label{bfa64}
\end{equation}
where $S_{2}=\int d^{4}xb$, which requires that $\Delta $ given in (\ref
{bfa63}) should be a $s$-coboundary modulo $d$ in order to have solutions
for the second-order deformation. This happens if and only if the function $%
U\left( \varphi \right) $ actually reduces to a constant, which we fix for
convenience to be equal to the unit 
\begin{equation}
U\left( \varphi \right) =1.  \label{ufi}
\end{equation}
By virtue of this result, we further infer that 
\begin{equation}
\frac{1}{2}\Delta =s\left( -\Phi \bar{\Phi}A^{\mu }A_{\mu }\right) +\partial
_{\mu }\left( 2\Phi \bar{\Phi}A^{\mu }\eta \right) ,  \label{bfa65}
\end{equation}
such that $b=\Phi \bar{\Phi}A^{\mu }A_{\mu }$, from which we can write down
the second-order deformation of the solution to the master equation under
the form 
\begin{equation}
S_{2}=\int d^{4}x\Phi \bar{\Phi}A^{\mu }A_{\mu }.  \label{bfa66}
\end{equation}
In the meantime, replacing (\ref{ufi}) back in (\ref{s1}) we determine that
the part from the first-order deformation that describes the cross-couplings
between the BF-field sector and the matter theory becomes 
\begin{equation}
i\left( A_{\mu }\left( \bar{\Phi}\partial ^{\mu }\Phi -\Phi \partial ^{\mu }%
\bar{\Phi}\right) +\left( \Phi ^{*}\Phi -\bar{\Phi}^{*}\bar{\Phi}\right)
\eta \right) .  \label{bfa66a}
\end{equation}
As $\left( S_{1},S_{2}\right) =0$, the third-order deformation equation
holds if we take $S_{3}=0$. All the other higher-order equations are then
satisfied with the choice $S_{4}=S_{5}=\cdots =0$.

Putting together the general results discussed at case (I) in the previous
section together with the above ones, we infer that the Lagrangian action of
the interacting model has the expression 
\begin{eqnarray}
&&\tilde{S}^{\prime }\left[ A^{\mu },H^{\mu },\varphi ,B^{\mu \nu },\Phi ,%
\bar{\Phi}\right] =\int d^{4}x\left( H_{\mu }\left( \partial ^{\mu }\varphi
-gWA^{\mu }\right) \right.  \nonumber \\
&&\left. +\frac{1}{2}B^{\mu \nu }\partial _{\left[ \mu \right. }A_{\left.
\nu \right] }+\left( \tilde{D}_{\mu }\Phi \right) \left( \overline{\tilde{D}%
^{\mu }\Phi }\right) -\mu ^{2}\Phi \bar{\Phi}-V\left( \Phi \bar{\Phi}\right)
\right) ,  \label{bfa69}
\end{eqnarray}
where the covariant derivative for the complex scalar field is 
\begin{equation}
\tilde{D}_{\mu }=\partial _{\mu }-igA_{\mu }.  \label{bfa68}
\end{equation}
The gauge transformations of the matter fields read as 
\begin{equation}
\bar{\delta}_{\epsilon }\Phi =ig\Phi \epsilon ,\;\bar{\delta}_{\epsilon }%
\bar{\Phi}=-ig\bar{\Phi}\epsilon ,  \label{bfa70}
\end{equation}
and can be obtained by directly gauging the initial rigid symmetry (\ref
{bfa61}). The gauge transformations of the BF field spectrum have been
inferred in the above and are expressed by the appropriate formulas in (\ref
{bfa49}--\ref{bfa50}) where we must take into account the result (\ref{ufi}%
), which further implies that $\delta U/\delta \varphi =0$. As we have
anticipated in Section 3 at the case (I.2), the non-invariance of the matter
current under the deformed gauge transformations of the matter fields at the
first order in the coupling constant restricts the form of the function $%
U\left( \varphi \right) $ in order to ensure the consistency of the
first-order deformation. Thus, the cross-interactions that can be added to
the BF model under study plus a complex scalar theory cannot involve an
arbitrary `background' of the undifferentiated scalar field from the BF
sector. In this light, the complex scalar field still gains gauge
transformations (see (\ref{bfa70})), but they merely reduce to the gauge
version of the original rigid one-parameter symmetries, the presence of an
arbitrary `background' being forbidden. Along the same line, we see from the
second relation in (\ref{bfa49}) with $\delta U/\delta \varphi $ replaced by
zero that the gauge transformations of the one-form $H^{\mu }$ cannot depend
on the matter fields 
\begin{equation}
\bar{\delta}_{\epsilon }H^{\mu }=2D_{\nu }\epsilon ^{\mu \nu }-g\frac{\delta
W}{\delta \varphi }H^{\mu }\epsilon ,  \label{bfa68a}
\end{equation}
where $D_{\nu }$ is defined in (\ref{bfa49a}).

\subsection{Massive spin 3/2 field}

In this situation, the role of the matter fields is played by a massive spin
3/2 field $y^{i}=\left( \psi _{\mu }^{A},\bar{\psi}_{\mu A}\right) $,
described by the Lagrangian 
\begin{equation}
\mathcal{L}_{0}^{\mathrm{matt}}=-\frac{1}{2}\varepsilon ^{\mu \nu \rho
\lambda }\bar{\psi}_{\mu A}\left( \gamma _{5}\gamma _{\nu }\right)
_{\;\;B}^{A}\partial _{\rho }\psi _{\lambda }^{B}+\frac{m}{2}\bar{\psi}_{\mu
A}\left( \sigma ^{\mu \nu }\right) _{\;\;B}^{A}\psi _{\nu }^{B},
\label{bfa70aa}
\end{equation}
where 
\begin{equation}
\sigma ^{\mu \nu }=\frac{1}{2}\left( \gamma ^{\mu }\gamma ^{\nu }-\gamma
^{\nu }\gamma ^{\mu }\right) .  \label{bfa71}
\end{equation}
The Koszul-Tate differential and the exterior longitudinal derivative
associated with this free model act on the matter fields and their
antifields $y_{i}^{*}=\left( \psi _{A}^{*\mu },\bar{\psi}^{*\mu A}\right) $
via the relations 
\begin{equation}
\delta \bar{\psi}_{\mu A}=\delta \psi _{\mu }^{A}=0,  \label{bfa72}
\end{equation}
\begin{equation}
\delta \bar{\psi}^{*\mu A}\equiv -\frac{\delta ^{L}\mathcal{L}_{0}^{\mathrm{%
matt}}}{\delta \bar{\psi}_{\mu A}}=\frac{1}{2}\varepsilon ^{\mu \nu \rho
\lambda }\left( \gamma _{5}\gamma _{\nu }\right) _{\;\;B}^{A}\partial _{\rho
}\psi _{\lambda }^{B}-\frac{m}{2}\left( \sigma ^{\mu \nu }\right)
_{\;\;B}^{A}\psi _{\nu }^{B},  \label{bfa73}
\end{equation}
\begin{equation}
\delta \psi _{A}^{*\mu }\equiv -\frac{\delta ^{L}\mathcal{L}_{0}^{\mathrm{%
matt}}}{\delta \psi _{\mu }^{A}}=-\frac{1}{2}\varepsilon ^{\mu \nu \rho
\lambda }\left( \gamma _{5}\gamma _{\nu }\right) _{\;\;A}^{B}\partial _{\rho
}\bar{\psi}_{\lambda B}-\frac{m}{2}\left( \sigma ^{\mu \nu }\right)
_{\;\;A}^{B}\bar{\psi}_{\nu B},  \label{bfa74}
\end{equation}
\begin{equation}
\gamma \bar{\psi}_{\mu A}=\gamma \psi _{\mu }^{A}=\gamma \bar{\psi}^{*\mu
A}=\gamma \psi _{A}^{*\mu }=0.  \label{bfa75}
\end{equation}
The functions defining the field equations are independent, like for the
previous example, and, moreover, they are linear in the fields and
first-order in their spacetime derivatives, such that the massive spin 3/2
fields are indeed described by a normal theory without gauge invariance, of
Cauchy order one, so the results inferred in the theoretical part of this
paper apply to this model. The actions of $\delta $ and $\gamma $ on the
BF-field spectrum can be recovered from the definitions (\ref{bfa15}--\ref
{bfa23}). If we multiply (\ref{bfa73}) from the left by $\bar{\psi}_{\mu A}$%
, (\ref{bfa74}) from the right by $\psi _{\mu }^{A}$, and subtract the
resulting relations, we obtain that 
\begin{equation}
\delta \left( -\psi _{A}^{*\mu }\psi _{\mu }^{A}+\bar{\psi}_{\mu A}\bar{\psi}%
^{*\mu A}\right) =\partial _{\mu }\left( -\frac{1}{2}\varepsilon ^{\mu \nu
\rho \lambda }\bar{\psi}_{\rho A}\left( \gamma _{5}\gamma _{\nu }\right)
_{\;\;B}^{A}\psi _{\lambda }^{B}\right) ,  \label{bfa76}
\end{equation}
which is related to the conservation of the non-trivial current 
\begin{equation}
j^{\mu }=-\frac{1}{2}\varepsilon ^{\mu \nu \rho \lambda }\bar{\psi}_{\nu
A}\left( \gamma _{5}\gamma _{\rho }\right) _{\;\;B}^{A}\psi _{\lambda }^{B},
\label{bfa77}
\end{equation}
associated with the bosonic global one-parameter invariance of the massive
spin 3/2 theory 
\begin{equation}
\Delta \psi _{\mu }^{A}=-\psi _{\mu }^{A}\xi ,\;\Delta \bar{\psi}_{\mu A}=%
\bar{\psi}_{\mu A}\xi .  \label{bfa78}
\end{equation}
If we compare (\ref{bfa78}) with (\ref{bfa40}), we find that 
\begin{equation}
T^{i}=\left( -\psi _{\mu }^{A},\bar{\psi}_{\mu A}\right) .  \label{bfa79}
\end{equation}

The first-order deformation is given by (\ref{bfa44}), where 
\begin{eqnarray}
&&A_{\mu }U\left( \varphi \right) j^{\mu }\left( \left[ y^{j}\right] \right)
+\left( H_{\mu }^{*}\frac{\delta U}{\delta \varphi }j^{\mu }\left( \left[
y^{j}\right] \right) +y_{i}^{*}U\left( \varphi \right) T^{i}\left( \left[
y^{j}\right] \right) \right) \eta \rightarrow  \nonumber \\
&&-\frac{1}{2}\varepsilon ^{\mu \nu \rho \lambda }U\left( \varphi \right) 
\bar{\psi}_{\mu A}\left( \gamma _{5}\gamma _{\nu }\right)
_{\;\;B}^{A}A_{\rho }\psi _{\lambda }^{B}+\left( -\frac{1}{2}\varepsilon
^{\mu \nu \rho \lambda }H_{\mu }^{*}\frac{\delta U}{\delta \varphi }\bar{\psi%
}_{\nu A}\left( \gamma _{5}\gamma _{\rho }\right) _{\;\;B}^{A}\psi _{\lambda
}^{B}\right.  \nonumber \\
&&\left. +U\left( \varphi \right) \left( -\psi _{A}^{*\mu }\psi _{\mu }^{A}+%
\bar{\psi}^{*\mu A}\bar{\psi}_{\mu A}\right) \right) \eta .  \label{s1model2}
\end{eqnarray}
By direct computation, we get that 
\begin{equation}
\left( S_{1},S_{1}\right) =0,  \label{bfa80}
\end{equation}
which is normal since the current (\ref{bfa77}) is invariant under the gauge
version of the transformations (\ref{bfa78}) (formula (\ref{bfa46}) is now
valid). Thus, this model subscribes to the situation (I.1) discussed in the
previous section. In this case we have shown that all the higher-order
deformations can be taken equal to zero, $S_{k}=0$, $k\geq 2$, such that the
Lagrangian action of the interacting model becomes 
\begin{eqnarray}
&&\tilde{S}\left[ A^{\mu },H^{\mu },\varphi ,B^{\mu \nu },\psi _{\mu }^{A},%
\bar{\psi}_{\mu A}\right] =  \nonumber \\
&&\int d^{4}x\left( H_{\mu }\left( \partial ^{\mu }\varphi -gW\left( \varphi
\right) A^{\mu }\right) +\frac{1}{2}B^{\mu \nu }\partial _{\left[ \mu
\right. }A_{\left. \nu \right] }\right.  \nonumber \\
&&\left. -\frac{1}{2}\varepsilon ^{\mu \nu \rho \lambda }\bar{\psi}_{\mu
A}\left( \gamma _{5}\gamma _{\nu }\right) _{\;\;B}^{A}\tilde{D}_{\rho
}^{\prime }\psi _{\lambda }^{B}+\frac{m}{2}\bar{\psi}_{\mu A}\left( \sigma
^{\mu \nu }\right) _{\;\;B}^{A}\psi _{\nu }^{B}\right) ,  \label{bfa81}
\end{eqnarray}
where 
\begin{equation}
\tilde{D}_{\rho }^{\prime }=\partial _{\rho }+gU\left( \varphi \right)
A_{\rho }.  \label{bfa82}
\end{equation}
The gauge transformations of the massive spin 3/2 field result from the
multiplication of the gauge version of the global transformations (\ref
{bfa78}) with the arbitrary function $U\left( \varphi \right) $%
\begin{equation}
\bar{\delta}_{\epsilon }\psi _{\mu }^{A}=-gU\left( \varphi \right) \psi
_{\mu }^{A}\epsilon ,\;\bar{\delta}_{\epsilon }\bar{\psi}_{\mu A}=gU\left(
\varphi \right) \bar{\psi}_{\mu A}\epsilon .  \label{bfa83}
\end{equation}
Moreover, the one-form $H_{\mu }$ gains gauge transformations that actually
depend on the matter fields 
\begin{equation}
\bar{\delta}_{\epsilon }H^{\mu }=2D_{\nu }\epsilon ^{\mu \nu }-g\left( \frac{%
\delta W}{\delta \varphi }H^{\mu }+\frac{1}{2}\varepsilon ^{\mu \nu \rho
\lambda }\frac{\delta U}{\delta \varphi }\bar{\psi}_{\nu A}\left( \gamma
_{5}\gamma _{\rho }\right) _{\;\;B}^{A}\psi _{\lambda }^{B}\right) \epsilon .
\label{bga83a}
\end{equation}
This completes the analysis of the second example.

\section{Conclusion}

In this paper we have investigated the local and manifestly covariant
Lagrangian interactions that can be added to a ``free'' theory describing a
BF-like model and a matter theory in four dimensions. Our treatment is based
on the deformation of the solution to the master equation. The first-order
deformation is computed by means of the local BRST cohomology at ghost
number zero. Its consistency reveals two cases. One of them brings no
cross-couplings between the BF and the matter fields, but merely
self-interactions among the BF ones, and ``cuts'' the deformation algorithm
to order one in the coupling constant. In the complementary situation we
find that the existence of non-trivial cross-interactions between the matter
and the BF fields requires that the matter theory is invariant under a
bosonic rigid one-parameter symmetry, which leads to a non-trivial conserved
current. The appearance of higher-order deformations is dictated by the
behavior of this current under the gauge version of the global invariance.
Thus, if the current is gauge invariant, then all the deformations of order
two and higher can be set equal to zero. In the opposite situation, at least
the second-order deformation of the solution to the master equation is
non-vanishing, but there can appear other non-trivial deformations as well.
The resulting coupled Lagrangian system exhibits many interesting features.
The interaction vertices are of two types: one involves only the BF field
sector and stops at the first order in the coupling constant, while the
other couples the matter fields to the BF ones through the conserved current
at the first order via a generalized minimal coupling of the form $U\left(
\varphi \right) j^{\mu }\left( \left[ y^{i}\right] \right) A_{\mu }$, and
might be non-trivial also at higher-order levels. The gauge transformations
change with respect to the ``free'' theory. In this context, the matter
fields are endowed at the first order in the coupling constant with the
gauge version of the purely matter global transformation multiplied by the
`background' potential $U\left( \varphi \right) $. Related to the new gauge
transformations of the BF field sector, we note that the gauge
transformation of the original $U\left( 1\right) $-abelian vector is not
modified. By contrast, the gauge transformations of the other one-form
present in the BF field spectrum are enriched with first-order (in the
interaction parameter) terms, of which one is linear in the matter current
and involves the derivative of the `background' potential, while the others
depend only on the BF fields. The gauge transformations of the remaining BF
fields (one scalar and one two-form) are also modified at the first order in
the coupling constant, but only through terms that involve the BF field
spectrum. The deformed gauge generators preserve the original second-stage
reducibility, but the reducibility relations of the interacting model only
hold on-shell. Moreover, the initial abelian gauge algebra is deformed into
an open one. In the case where the matter current is not invariant under the
gauge version of the initial one-parameter global symmetry we expect that
the consistency of the deformed solution to the master equation at higher
orders in the coupling constant may impose certain restrictions on the
`background' potential. The theoretical part of the paper has been
exemplified on two kinds of matter fields, namely, the massive complex
scalar theory and the massive spin 3/2 model. In the first case, the
conserved current is not invariant under the gauge version of the global
symmetry, and consequently there appear non-trivial interactions between the
complex scalar field and the BF theory, but at the second-order in the
constant coupling only. Meanwhile, the `background' cannot effectively
depend on the scalar field from the BF-sector; it is restricted to be
constant. In the other situation, the conserved current is gauge invariant,
so the deformation procedure stops after the first order and the
`background' is fully manifested.

\section*{Acknowledgment}

This work has been supported by the type $A_T$ grant 33547/2003, code
302/2003, from the Romanian Council for Academic Scientific Research
(C.N.C.S.I.S.) and the Romanian Ministry of Education, Research and
Youth (M.E.C.T.). The authors thank
the Referee for his/her valuable comments and suggestions.


\begin{thebibliography}{99}
\bibitem{1a}  E. S. Fradkin, G. A. Vilkovisky, Phys. Lett. \textbf{B55}
(1975) 224

\bibitem{1b}  I. A. Batalin, G. A. Vilkovisky, Phys. Lett. \textbf{B69}
(1977) 309

\bibitem{1c}  E. S. Fradkin, T. E. Fradkina, Phys. Lett. \textbf{B72} (1978)
343

\bibitem{1d}  I. A. Batalin, G. A. Vilkovisky, Phys. Lett. \textbf{B102}
(1981) 27

\bibitem{1e}  I. A. Batalin, E. S. Fradkin, Phys. Lett. \textbf{B122} (1983)
157

\bibitem{1f}  I. A. Batalin, G. A. Vilkovisky, Phys. Rev. \textbf{D28}
(1983) 2567

\bibitem{1g}  I. A. Batalin, G. A. Vilkovisky, J. Math. Phys. \textbf{26}
(1985) 172

\bibitem{1h}  M. Henneaux, Phys. Rep. \textbf{126} (1985) 1

\bibitem{1i}  A. D. Browning, D. Mc Mullan, J. Math. Phys. \textbf{28}
(1987) 438

\bibitem{1j}  M. Dubois-Violette, Ann. Inst. Fourier \textbf{37} (1987) 45

\bibitem{1k}  D. Mc Mullan, J. Math. Phys. \textbf{28} (1987) 428

\bibitem{1l}  M. Henneaux, C. Teitelboim, Commun. Math. Phys. \textbf{115}
(1988) 213

\bibitem{1m}  J. D. Stasheff, Bull. Amer. Math. Soc. \textbf{19} (1988) 287

\bibitem{1n}  J. Fisch, M. Henneaux, J. D. Stasheff, C. Teitelboim, Commun.
Math. Phys. \textbf{120} (1989) 379

\bibitem{1p}  M. Henneaux, Nucl. Phys. \textbf{B} (Proc. Suppl) \textbf{18A}
(1990) 47

\bibitem{1q}  M. Henneaux, C. Teitelboim, Quantization of Gauge Systems
(Princeton University Press, Princeton, New Jersey 1992)

\bibitem{4a}  B. Voronov, I. V. Tyutin, Theor. Math. Phys. \textbf{50}
(1982) 218

\bibitem{4b}  B. Voronov, I. V. Tyutin, Theor. Math. Phys. \textbf{52}
(1982) 628

\bibitem{4c}  J. Gomis, S. Weinberg, Nucl. Phys. \textbf{B469} (1996) 473,
hep-th/9510087

\bibitem{4d}  S. Weinberg, The Quantum Theory of Fields (Cambridge
University Press, Cambridge 1996)

\bibitem{5}  O. Piguet, S. P. Sorella, Algebraic Renormalization:
Perturbative Renormalization, Symmetries and Anomalies (Lecture Notes in
Physics, Springer Verlag, Berlin, Vol. \textbf{28}, 1995)

\bibitem{6c}  P. S. Howe, V. Lindstr\H{o}m, P. White, Phys. Lett. \textbf{%
B246} (1990) 130

\bibitem{6b}  W. Troost, P. van Nieuwenhuizen, A. van Proeyen, Nucl. Phys. 
\textbf{B333} (1990) 727

\bibitem{6a}  G. Barnich, M. Henneaux, Phys. Rev. Lett. \textbf{72} (1994)
1588, hep-th/9312206

\bibitem{6d}  G. Barnich, Mod. Phys. Lett.\textbf{A9} (1994) 665,
hep-th/9310167

\bibitem{6}  G. Barnich, Phys. Lett. \textbf{B419} (1998) 211, hep-th/9710162

\bibitem{7}  F. Brandt, M. Henneaux, A. Wilch, Phys. Lett. \textbf{B387}
(1996) 320, hep-th/9606172

\bibitem{7a}  R. Arnowitt, S. Deser, Nucl. Phys. \textbf{49} (1963) 133

\bibitem{7b}  J. Fang, C. Fronsdal, J. Math. Phys. \textbf{20} (1979) 2264

\bibitem{7c}  F. A. Berends, G. H. Burgers, H. Van Dam, Z. Phys. \textbf{C24}
(1984) 247

\bibitem{7d}  F. A. Berends, G. H. Burgers, H. Van Dam, Nucl. Phys. \textbf{%
B260} (1985) 295

\bibitem{7e}  A. K. H. Bengtsson, Phys. Rev. \textbf{D32} (1985) 2031

\bibitem{8a}  G. Barnich, M. Henneaux, Phys. Lett. \textbf{B311} (1993) 123,
hep-th/9304057

\bibitem{8b}  J. D. Stasheff, preprint q-alg/9702012

\bibitem{8c}  J. D. Stasheff, preprint hep-th/9712157

\bibitem{17and5}  M. Henneaux, Contemp. Math. \textbf{219} (1998) 93,
hep-th/9712226

\bibitem{8d}  J. A. Garcia, B. Knaepen, Phys. Lett. \textbf{B441} (1998)
198, hep-th/9807016

\bibitem{9a}  E. Cremmer, B. Julia, J. Scherk, Phys. Lett. \textbf{B76}
(1978) 409

\bibitem{9b}  R. Wald, Phys. Rev. \textbf{D33} (1986) 3613

\bibitem{9c}  G. Barnich, M. Henneaux, R. Tatar, Int. J. Mod. Phys. \textbf{%
D3} (1994) 139, hep-th/9307155

\bibitem{9d}  G. Barnich, F. Brandt, M. Henneaux, Commun. Math. Phys. 
\textbf{174} (1995) 93, hep-th/9405194

\bibitem{9e}  G. Barnich, F. Brandt, M. Henneaux, Nucl. Phys. \textbf{B455}
(1995) 357, hep-th/9505173

\bibitem{9f}  M. Henneaux, Phys. Lett. \textbf{B368} (1996) 83,
hep-th/9511145

\bibitem{9g}  M. Henneaux, B. Knaepen, Phys. Rev. \textbf{D56} (1997) 6076,
hep-th/9706119

\bibitem{9h}  M. Henneaux, B. Knaepen, C. Schomblond, Lett. Math. Phys. 
\textbf{42} (1997) 337, hep-th/9702042

\bibitem{9i}  M. Henneaux, B. Knaepen, C. Schomblond, Commun. Math. Phys. 
\textbf{186} (1997) 137, hep-th/9606181

\bibitem{9j}  M. Henneaux, V. E. R. Lemes, C. A. G. Sasaki, S. P. Sorella,
O. S. Ventura, I. C. Q. Vilar, Phys. Lett. \textbf{B410} (1997) 195,
hep-th/9707129

\bibitem{9k}  F. Brandt, Ann. Phys. (N.Y.) \textbf{259} (1997) 253,
hep-th/9609192

\bibitem{9l}  X. Bekaert, M. Henneaux, Int. J. Theor. Phys. \textbf{38}
(1999) 1161, hep-th/9806062

\bibitem{9m}  X. Bekaert, M. Henneaux, A. Sevrin, Phys. Lett. \textbf{B468}
(1999) 228, hep-th/9909094

\bibitem{9n}  C. Bizdadea, M. G. Mocioac\u {a}, S. O. Saliu, Phys. Lett. 
\textbf{B459} (1999) 145, hep-th/0003193

\bibitem{9p}  X. Bekaert, M. Henneaux, A. Sevrin, Nucl. Phys. \textbf{B}
(Proc. Suppl.) \textbf{88} (2000) 27, hep-th/9912077

\bibitem{9o}  K. I. Izawa, Prog. Theor. Phys. \textbf{103} (2000) 225,
hep-th/9910133

\bibitem{9q}  C. Bizdadea, L. Saliu, S. O. Saliu, Phys. Scripta \textbf{61}
(2000) 307, hep-th/0008022

\bibitem{9r}  N. Ikeda, JHEP \textbf{0011} (2000) 009, hep-th/0010096

\bibitem{9s}  N. Ikeda, JHEP \textbf{0107} (2001) 037, hep-th/0105286

\bibitem{9t}  X. Bekaert, M. Henneaux, A. Sevrin, Commun. Math. Phys. 
\textbf{224} (2001) 683, hep-th/0004049

\bibitem{9u}  N. Boulanger, T. Damour, L. Gualtieri, M. Henneaux, Nucl.
Phys. \textbf{B597} (2001) 127, hep-th/0007220

\bibitem{9v}  C. Bizdadea, E. M. Cioroianu, I. Negru, S. O. Saliu, Annalen
Phys. \textbf{10} (2001) 415

\bibitem{9x}  C. Bizdadea, M. T. Miaut\u {a}, S. O. Saliu, Acta Phys. Polon. 
\textbf{B32} (2001) 1225

\bibitem{9y}  C. Bizdadea, E. M. Cioroianu, M. T. Miaut\u {a}, I. Negru, S.
O. Saliu, Annalen Phys. \textbf{10} (2001) 921, hep-th/0201073

\bibitem{9w}  N. Boulanger, Fortsch. Phys. \textbf{50} (2002) 858,
hep-th/0111216

\bibitem{9z}  X. Bekaert, N. Boulanger, M. Henneaux, Phys. Rev. \textbf{D67}
(2003) 044010, hep-th/0210278

\bibitem{9zb}  C. Bizdadea, E. M. Cioroianu, I. Negru, S. O. Saliu, Eur.
Phys. J. \textbf{C27} (2003) 457, hep-th/0211158

\bibitem{9za}  F. Brandt, JHEP \textbf{0304} (2003) 035, hep-th/0212070

\bibitem{12}  D. Birmingham, M. Blau, M. Rakowski, G. Thompson, Phys. Rep. 
\textbf{209} (1991) 129

\bibitem{psm1}  N. Ikeda, Ann. Phys. (N.Y.) \textbf{235} (1994) 435, hep-th/9312059

\bibitem{psm1a}  T. Strobl, Phys. Rev. \textbf{D50} (1994) 7346,
hep-th/9403121

\bibitem{stroblspec}  P. Schaller, T. Strobl, Mod. Phys. Lett. \textbf{A9}
(1994) 3129, hep-th/9405110

\bibitem{psm2}  A. Yu. Alekseev, P. Schaller, T. Strobl, Phys. Rev. \textbf{%
D52} (1995) 7146, hep-th/9505012

\bibitem{psm4}  T. Kloesch, T. Strobl, Class. Quantum Grav. \textbf{13}
(1996) 965, gr-qc/9508020; Erratum-ibid. \textbf{14} (1997) 825

\bibitem{psm5}  T. Kloesch, T. Strobl, Class. Quantum Grav. \textbf{13}
(1996) 2395, gr-qc/9511081

\bibitem{psm6}  T. Kloesch, T. Strobl, Class. Quantum Grav. \textbf{14}
(1997) 1689, hep-th/9607226

\bibitem{psm3a}  A. S. Cattaneo, G. Felder, Commun. Math. Phys. \textbf{212}
(2000) 591, math.QA/9902090

\bibitem{psm3}  T. Strobl, Gravity in Two Spacetime Dimensions, Habilitation
thesis RWTH Aachen, May 1999, hep-th/0011240

\bibitem{psmn}  A. S. Cattaneo, G. Felder, Mod. Phys. Lett. \textbf{A16}
(2001) 179, hep-th/0102208

\bibitem{gen1}  G. Barnich, F. Brandt, M. Henneaux, Commun. Math. Phys. 
\textbf{174} (1995) 57, hep-th/9405109

\bibitem{gen2}  G. Barnich, F. Brandt, M. Henneaux, Phys. Rep. \textbf{338}
(2000) 439, hep-th/0002245
\end{thebibliography}
\end{document}